\documentclass[twocolumn,prx,superscriptaddress, longbibliography]{revtex4-1}
\usepackage[colorlinks=true,urlcolor=blue,citecolor=blue,linkcolor=blue]{hyperref}
\usepackage{amssymb}
\usepackage{bm}
\usepackage{amsmath}
\usepackage{graphicx}
\usepackage{epstopdf}
\usepackage{subfig}
\usepackage{epsfig}
\usepackage{amsfonts}
\usepackage{mathrsfs}
\usepackage{mathdots}
\usepackage{float}
\usepackage{tikz}
\usepackage{txfonts}
\usepackage[normalem]{ulem}

\newcommand{\Eq}[1]{Eq.~(\ref{#1})}

\newcommand{\tn}[2]{\begin{minipage}[c]{#1\linewidth}
\centering
\includegraphics[page=#2]{TN.pdf}
\end{minipage}}
\def\Tr{{\mathrm{Tr}}}

\newcommand{\dmax}{\mathcal{D}_{\mathrm{max}}}

\begin{document}
\title{Unsupervised Generative Modeling Using Matrix Product States}
\author{Zhao-Yu Han}
\thanks{These two authors contributed equally}
\affiliation{School of Physics, Peking University, Beijing 100871, China}
\author{Jun Wang}
\thanks{These two authors contributed equally}
\affiliation{School of Physics, Peking University, Beijing 100871, China}
\author{Heng Fan}
\affiliation{Institute of Physics, Chinese Academy of Sciences, Beijing 100190, China}
\author{Lei Wang}
\email{wanglei@iphy.ac.cn}
\affiliation{Institute of Physics, Chinese Academy of Sciences, Beijing 100190, China}
\author{Pan Zhang}
\email{panzhang@itp.ac.cn}
\affiliation{Institute of Theoretical Physics, Chinese Academy of Sciences, Beijing 100190, China}
\begin{abstract}
Generative modeling, 
which learns joint probability distribution from data and generates samples according to it, is an important task in machine learning and artificial intelligence. 
Inspired by probabilistic interpretation of quantum physics, we propose a generative model using \textit{matrix product states}, which is a tensor network originally proposed for describing (particularly one-dimensional) entangled quantum states. 
Our model enjoys efficient learning analogous to the \textit{density matrix renormalization group} method, which allows dynamically adjusting dimensions of the tensors and offers an efficient direct sampling approach for generative tasks.
We apply our method to generative modeling of several standard datasets including the 
Bars and Stripes
random binary patterns and the 
MNIST
handwritten digits to illustrate the abilities, features and drawbacks of our model over popular generative models such as Hopfield model, Boltzmann machines and generative adversarial networks. Our work sheds light on many interesting directions of future exploration on the development of quantum-inspired algorithms for unsupervised machine learning, which are promisingly possible to be realized on quantum devices.

%
\end{abstract}
\maketitle
%

\section{Introduction}
Generative modeling, a typical example of unsupervised learning that makes use of huge amount of unlabeled data, lies in the heart of rapid development of modern machine learning techniques~\cite{lecun2015deep}. Different from discriminative tasks such as pattern recognition, the goal of generative modeling is to model the probability distribution of data and thus be able to generate \emph{new} samples according to the distribution. 
At the research frontier of generative modeling, it is used for finding good data representation and dealing with tasks with missing data. Popular generative machine learning models include Boltzmann Machines (BM)~\cite{Ackley1985,smolensky1986foundations} and their generalizations~\cite{salakhutdinov2015learning}, variational autoencoders (VAE)~\cite{VAE}, autoregressive models~\cite{NADE,pixelRNN}, normalizing flows~\cite{NICE,realNVP,rezende2015variational}, and generative adversarial networks (GAN)~\cite{GAN}. 
For generative model design, one tries to balance the representational power and  efficiency of learning and sampling.

There is a long history of the interplay between generative modeling and statistical physics. Some celebrated models, such as Hopfield model~\cite{Hopfield1982} and Boltzmann machine~\cite{Ackley1985, smolensky1986foundations}, are closely related to the Ising model and its inverse version which learns couplings in the model based on given training configurations~\cite{kappen1998boltzmann,Roudi2009a}. 

The task of generative modeling also shares similarities with quantum physics in the sense that both of them try to model probability distributions in an immense space. 
Precisely speaking, it is the wavefunctions that are modeled in quantum physics, and probability distributions are given by their squared norm according to Born's statistical interpretation. 
Modeling probability distributions in this way is fundamentally different from the traditional statistical physics perspective. Hence we may refer probability models which exploit quantum state representations as ``{\it Born Machines}''. Various ansatz have been developed to express quantum states, such as the variational Monte Carlo~\cite{QMC-RevModPhys.73.33}, the tensor network (TN) states and recently artificial neural networks~\cite{Carleo2017} . In fact physical systems like quantum circuits are also promising candidates for implementing Born Machines.

In the past decades, tensor network states and algorithms have been shown to be an incredibly potent tool set for modeling many-body quantum states ~\cite{Schollwock2011, ORUS2014117}. The success of TN description can be theoretically justified from a quantum information perspective~\cite{area, Eisert2010}. In parallel to quantum physics applications, tensor decomposition and tensor networks have also been applied in a broader context by the machine learning community for feature extraction, dimensionality reduction and analyzing the expressibility of deep neural networks~\cite{BenguaPT15, novikov2014putting, novikov2015tensorizing, CohenSS15a, cichocki2016tensor, 2017arXiv170401552L}. 

In particular, matrix product state (MPS) is a kind of TN where the tensors are arranged in a one-dimensional geometry~\cite{MPS}. The same representation is referred as tensor train decomposition in the applied math community~\cite{tensortrain}. Despite its simple structure, MPS can represent a large number of quantum states extremely well. MPS representation of ground states has been proven to be efficient for one-dimensional gapped local Hamiltonian~\cite{landau2013polynomial}. In practice, optimization schemes for MPS such as density-matrix renormalization group (DMRG)~\cite{DMRG} have been successful even for some quantum systems in higher dimension~\cite{2DDMRG}. 
Some recent works extended the application of MPS to machine learning tasks like pattern recognition ~\cite{schwab}, classification~\cite{ novikov2016exponential} and language modeling~\cite{gallego2017physical}. Efforts also drew connection between Boltzmann Machines and tensor networks~\cite{equiv-PhysRevB.97.085104}.

In this paper, building on the connection between unsupervised  generative modeling and quantum physics, we employ MPS as a model to learn probability distribution of given data with an algorithm which resembles DMRG~\cite{DMRG}. 
Compared with statistical-physics based models such as the Hopfield model~\cite{Hopfield1982} and the inverse Ising model, MPS exhibits much stronger ability of learning, which adaptively grows by increasing bond dimensions of the MPS. 
The MPS model also enjoys a direct sampling method~\cite{PhysRevB.85.165146} much more efficient than the Boltzmann machines, which require Markov Chain Monte Carlo (MCMC) process for data generation.
When compared with popular generative models such as GAN, our model offers a more efficient way to reconstruct and denoise from an initial (noisy) input using the direct sampling algorithm, as opposed to GAN where mapping a noisy image to its input is not straightforward.

The rest of the paper is organized as follow. In Sec.~\ref{sec:algorithm} we present our model, training algorithm and  direct sampling method. In Sec.~\ref{sec:expts} we apply our model to three datasets: Bars-and-stripes for a proof-of-principle demonstration, random binary patterns for capacity illustration and the MNIST handwritten digits for showing the generalization ability of the MPS model in unsupervised tasks such as reconstruction
of images. Finally, Sec~\ref{sec:discussion} discusses future prospects of the generative modeling using more general tensor networks and quantum circuits. 

\section{MPS for Unsupervised Learning \label{sec:algorithm}}
The goal of unsupervised generative modeling is to model the joint probability distribution of given data. With the trained model, one can then generate new samples from the learned probability distribution. Generative modeling finds wide applications such as dimensional reduction, feature detection, clustering and recommendation systems~\cite{Goodfellow-et-al-2016}. In this paper, we consider a data set $\mathcal{T}$ consisting of binary strings $\ensuremath{\boldsymbol{v}}\in \mathcal{V} = \{0, 1\}^{\otimes N}$, which are potentially repeated and can be mapped to basis vectors of a Hilbert space of dimension $2^N$. 

The probabilistic interpretation of quantum mechanics~\cite{MaxBorn} naturally suggests modeling data distribution with a quantum state. Suppose we encode the probability distribution into a quantum wavefunction $\Psi(\ensuremath{\boldsymbol{v}})$, measurement will collapse it and generate a result $\ensuremath{\boldsymbol{v}} = (v_1, v_2, \cdots, v_N)$ with a probability proportional to  $|\Psi(\ensuremath{\boldsymbol{v}})|^2$
Inspired by the generative aspects of quantum mechanics, we represent the model probability distribution by 
\begin{equation}
\mathbb{P}(\ensuremath{\boldsymbol{v}}) =
\frac{|\Psi(\ensuremath{\boldsymbol{v}})|^{2}}{Z}, 
\label{eq:quantum}
\end{equation}
where $Z = \sum_{\ensuremath{\boldsymbol{v}}\in\mathcal{V}}|\Psi(\ensuremath{\boldsymbol{v}})|^{2}$ is the normalization factor. We also refer it as the \emph{partition function} to draw an analogy with the energy based models~\cite{lecun2006tutorial}. In general the wavefunction $\Psi({\ensuremath{\boldsymbol{v}}})$  can be complex valued, but in this work we restrict it to be real valued. Representing probability density using square of a function was also put forward by former works~
\cite{schwab, Zhao-Jaeger, pmlr-v20-bailly11}. These approaches ensure the positivity of probability and naturally admit a quantum mechanical interpretation. 


\subsection{Matrix Product States}

Quantum physicists and chemists have developed many efficient classical representations of quantum wavefunctions. A number of these developed representations and algorithms can be adopted for efficient probabilistic modeling. Here, we parametrize the wave function 
using MPS:
\begin{align}
\Psi({v_1,v_2,\cdots,v_N})= 
\Tr \left(A^{(1)v_1}A^{(2)v_2}\cdots A^{(N)v_N}\right), 
\label{eq:MPS}
\end{align}
where each $A^{(k)v_{k}}$ is a $\mathcal{D}_{k-1}$ by $\mathcal{D}_k$ matrix, and $\mathcal{D}_0=\mathcal{D}_N$ is demanded to close the trace.  
For the the case considered here, there are $2\sum_{k=1}^{N}\mathcal{D}_{k-1}\mathcal{D}_{k}$ parameters on the right-hand-side of Eq.~(\ref{eq:MPS}). 
The representational power of MPS is related to Von Neumann entanglement entropy of the quantum state, which is defined as $S=-\Tr(\rho_{A} \ln \rho_{A})$. Here we divide the variables into two groups $\ensuremath{\boldsymbol{v}} = (\ensuremath{\boldsymbol{v}}_A, \ensuremath{\boldsymbol{v}}_B)$ and  $\rho_{A} = \sum_{\ensuremath{\boldsymbol{v}}_{B}} \Psi(\ensuremath{\boldsymbol{v}}_{A}, \ensuremath{\boldsymbol{v}}_{B}) \Psi( \ensuremath{\boldsymbol{v}}^{\prime}_{A}, \ensuremath{\boldsymbol{v}}_{B})$ is the reduced density matrix of a subsystem. The entanglement entropy sets a lower bound for the bond dimension at the division $S\le\ln(\mathcal{D}_{k})$. Any probability distribution of a $N$-bit system can be described by an MPS as long as its bond dimensions are free from any restriction. The inductive bias using MPS with limited bond dimensions comes from dropping off the minor components of entanglement spectrum. Therefore as the bond dimension increases, an MPS enhances its ability of parameterizing complicated functions. See \cite{Schollwock2011} and \cite{ORUS2014117} for recent reviews on MPS and its applications on quantum many-body systems. 

In practice, it is convenient to use MPS with $\mathcal{D}_{0}=\mathcal{D}_N=1$ and consequently reduce the left and right most matrices to vectors~\cite{DMRG}. In this case, Eq. (\ref{eq:MPS}) reads schematically
\begin{equation}
\tn{0.33}{1}
=
\tn{0.47}{2}.
\end{equation}
Here the blocks denote the tensors and the connected lines indicate tensor contraction over virtual indices. The dangling vertical bonds denote physical indices. We refer to~\cite{Schollwock2011, ORUS2014117} for an introduction to these graphical notations of TN. Henceforth, we shall present formulae with more intuitive graphical notations wherever possible. 

The MPS representation has gauge degrees of freedom, which allows one to restrict the tensors with canonical conditions. We remark that in our setting of generative modeling, the canonical form significantly benefits computing the {\it exact} partition function $Z$. More details about the canonical condition and the calculation of $Z$ can be found in Appendix~\ref{sec:canonical}.
  
\subsection{Learning MPS from Data}
Once the MPS form of wavefunction $\Psi(\ensuremath{\boldsymbol{v}})$ is chosen, learning can be achieved by adjusting parameters of the wave function such that the distribution represented by Born's rule Eq.~\eqref{eq:quantum} is as close as possible to the data distribution. A standard learning method is called {\it Maximum Likelihood Estimation} which defines a (negative) log-likelihood function and optimizes it by adjusting the parameters of the model.
In our case, the negative log-likelihood (NLL) is defined as
\begin{equation}
\mathcal {L} = -\frac{1}{|\mathcal{T}|}\sum_{\ensuremath{\boldsymbol{v}}\in \mathcal{T}}   \ln~\mathbb{P}(\ensuremath{\boldsymbol{v}}), 
\label{eq:NLL}
\end{equation}

where $|\mathcal{T}|$ denotes the size of the training set. Minimizing the NLL reduces the dissimilarity between the model probability distribution $\mathbb{P}(\ensuremath{\boldsymbol{v}})$ and the empirical distribution defined by the training set. It is well-known that minimizing $\mathcal{L}$ is equivalent to minimizing the Kullback-Leibler divergence between the two distributions~\cite{Kullback1951}. 

Armed with canonical form, we are able to differentiate the negative log-likelihood \eqref{eq:NLL} with respect to the components of an order-$4$ tensor $A^{(k,k+1)}$, which is obtained by contracting two adjacent tensors $A^{(k)}$ and $A^{(k+1)}$. The gradient reads
\begin{equation}
\label{eq:gradient:1}
\frac{\partial\mathcal{L}}{\partial A^{(k,k+1)w_kw_{k+1}}_{i_{k-1}i_{k+1}}} = \frac{ Z^{\prime}}{Z} -\frac{2}{|\mathcal{T}|} \sum_{\ensuremath{\boldsymbol{v}}\in \mathcal{T}}   \frac{\Psi^{\prime}(\ensuremath{\boldsymbol{v}} )}{\Psi(\ensuremath{\boldsymbol{v}})} , 
\end{equation}
where $\Psi^{\prime}(\ensuremath{\boldsymbol{v}} )$ denotes the derivative of the MPS with respect to the tensor element of $A^{(k,k+1)}$, and 
$Z^{\prime} = 2\sum_{\ensuremath{\boldsymbol{v}}\in \mathcal{V}} \Psi^{\prime}(\ensuremath{\boldsymbol{v}} )\Psi(\ensuremath{\boldsymbol{v}} ) $. 
Note that although $Z$ and $Z'$ involve summations over an exponentially large number of terms, they are tractable in the MPS model via efficient contraction schemes~\cite{Schollwock2011}. In particular, if the MPS is in the mixed-canonical form~\cite{Schollwock2011}, $Z'$ can be significantly simplified to $Z'=2 A^{(k,k+1)w_kw_{k+1}}_{i_{k-1}i_{k+1}}$. 
The calculation of the gradient, as well as variant techniques in gradient descent such as \textit{stochastic gradient descent}~(SGD) and adaptive learning rate, are detailed in Appendix~\ref{sec:dmrg}. After gradient descent, the merged order-4 tensor is decomposed into two order-3 tensors, and then the procedure is repeated for each pair of adjacent tensors. 

The derived algorithm is quite similar to the celebrated DMRG method with two-site update, which allows us to adjust dynamically the bond dimensions during the optimization and to allocate computational resources to the important bonds which represent essential features of data. 
However we emphasize that there are key differences between our algorithm and DMRG:
\begin{itemize}
\item{The loss function of classic DMRG method is usually the energy, while our loss function, the averaged NLL~\eqref{eq:NLL}, is a function of data. 
}
\item{With a huge amount of data, the landscape of the loss function is typically very complicated so that modern optimizers developed in the machine learning community, such as stochastic gradient descent and learning rate adapting techniques~\cite{kingma2014adam}, are important to our algorithm. Since the ultimate goal of learning is optimizing the performance on the test data, we do not really need to find the optimal parameters minimizing the loss on the training data. One usually stops training before reaching the actual minima to prevent overfitting.}
\item{Our algorithm is data-oriented. It is straightforward to parallelize over the samples since the operations applied to them are identical and independent. In fact, it is a common practice in modern deep learning framework to parallelize over this so-called "batch" dimension~\cite{Goodfellow-et-al-2016}. As a concrete example, the GPU implementation of our algorithm is at least $100$ times faster than the CPU implementation on the full MNIST dataset.}
\end{itemize}


%

\subsection{Generative Sampling\label{sec:Zipper}}
After training, samples can be generated independently according to Eq.~\eqref{eq:quantum}
.
In other popular generative models, especially the energy based model such as restricted Boltzmann machine (RBM)~\cite{smolensky1986foundations}, generating new samples is often accomplished by running MCMC from an initial configuration, due to the intractability of the partition function. In our model, one convenience is that the partition function can be exactly computed with complexity linear in system size. Our model enjoys
a direct sampling method which generates a sample bit by bit from one end of the MPS to the other~\cite{PhysRevB.85.165146}. 
The detailed generating process is as follow:

It starts from one end, say the $N$-th bit. One directly samples this bit from the 
marginal probability $\mathbb{P}(v_N)=\sum_{v_{1},v_{2},\ldots,v_{N-1}}\mathbb{P}(\ensuremath{\boldsymbol{v}}) $. It is clear that this can be easily performed if we have gauged all the tensors except $A^{(N)}$ to be left-canonical because $\mathbb{P}(v_N)=|\ensuremath{\boldsymbol{x}}^{v_{N}}|^{2}/Z$, where we define $x_{i_{N-1}}^{v_{N}}=A^{(N)v_{N}}_{i_{N-1}}$ and the normalization factor reads $Z=\sum_{v_{N}\in\{0,1\}}|\ensuremath{\boldsymbol{x}}^{v_{N}}|^2$. Given the value of the $N$-th bit, one can then move on to sample the $(N-1)$-th bit. More generally, given the bit values $v_k, v_{k+1},\cdots, v_N$, the $(k-1)$-th bit is sampled according to the conditional probability 
\begin{align}
\mathbb{P}(v_{k-1} | v_{k}, v_{k+1},\ldots,v_N) = \frac{\mathbb{P}(v_{k-1}, v_{k}, \ldots,v_N)}{\mathbb{P}(v_{k}, v_{k+1} \ldots,v_N)}. \label{eq:Pcondition}
\end{align}
As a result of the canonical condition, the marginal probability
can be simply expressed as 
\begin{align}
\mathbb{P}(v_{k}, v_{k+1}, \ldots,v_N)
&= | \ensuremath{\boldsymbol{x}}^{v_{k},v_{k+1},\ldots,v_{N}} |^{2}/Z. \label{eq:PvkN}
\end{align}
$x_{i_{k-1}}^{v_{k},v_{k+1},\ldots,v_{N}} = \sum_{i_{k},i_{k+1},\cdots, i_{N-1}} A^{(k)v_k}_{i_{k-1}i_{k}}A^{(k+1)v_{k+1}}_{i_{k}i_{k+1}}\cdots A^{(N)v_N}_{i_{N-1}}$ has been settled since the $k$-th bit is sampled. 
Schematically, its squared norm reads
\begin{equation}
| \ensuremath{\boldsymbol{x}}^{v_{k},v_{k+1},\ldots,v_{N}}|^{2} 
=
\tn{0.31}{15}.
\end{equation}
Multiplying the matrix $A^{(k-1)v_{k-1}}$ from the left, and calculating the squared norm of the resulting vector $x_{i_{k-2}}^{v_{k-1},v_{k},\ldots,v_{N}}=\sum_{i_{k-1}}A^{(k-1)v_{k-1}}_{i_{k-2}i_{k-1}}x_{i_{k-1}}^{v_{k},v_{k-1},\ldots,v_{N}}$, one obtains 
\begin{equation}
\mathbb{P}(v_{k-1}, v_{k}, ...,v_N) = |\ensuremath{\boldsymbol{x}}^{v_{k-1},v_{k},\ldots,v_{N}}|^2/Z. \label{eq:Pvk-1N}
\end{equation}
Combining (\ref{eq:PvkN}, \ref{eq:Pvk-1N}) one can compute the conditional probability (\ref{eq:Pcondition}) and sample the bit $v_{k-1}$ accordingly. In this way, all the bit values are successively drawn from the conditional probabilities given all the bits on the right. This procedure gives a sample strictly obeying the probability distribution of the MPS. 

This sampling approach is not limited to generating samples from scratch in a sequential order. It is also capable of inference tasks when part of the bits are given. 
In that case, the canonicalization trick may not help greatly if there is a segment of unknown bits sitting between given bits. Nevertheless, the marginal probabilities are still tractable because one can also contract ladder-shaped TN efficiently~\cite{Schollwock2011, ORUS2014117}. 
As what will be shown in Sec.~\ref{sec:expts}, given these flexibilities of the sampling approach, MPS-based probabilistic modeling can be applied to image reconstruction and denoising. 

\subsection{Features of the model and algorithms} \label{sec:features}
We highlight several salient features of the MPS generative model and compare it to other popular generative models. Most significantly, MPS has an explicit tractable probability density, while still allows efficient learning and inference.
For a system sized $N$, with prescribed maximal bond dimension $\dmax$, the complexity of training on a dataset of size $|\mathcal{T}|$ is $\mathcal{O}(|\mathcal{T}|  N \mathcal{D}_\mathrm{max}^{3})$. The scaling of generative sampling from a canonical MPS is $\mathcal{O}( N  \mathcal{D}_\mathrm{max}^{2})$ if all the bits to be sampled are connected to the boundaries, otherwise given some segments the conditional sampling scales as $\mathcal{O}( N  \mathcal{D}_\mathrm{max}^{3})$. 

\subsubsection{Theoretical Understanding of the Expressive Power}

The expressibility of MPS was intensively studied in the context of quantum physics. The bond dimensions of MPS put an upper bound on its ability of capturing entanglement entropy. These solid theoretical understandings of the representational power of MPS~\cite{Schollwock2011, ORUS2014117} makes it an appealing model for generative tasks.

Considering the success of MPS for quantum systems, we expect a polynomial scaling of the computational resources for datasets with short-range correlations. Treating dataset of two dimensional images using MPS is analogous to the application of DMRG to two dimensional quantum systems~\cite{2DDMRG}. Although in principle an exact representation of the image dataset may require exponentially large bond dimensions as the image resolution increases, 
at computationally affordable bond dimensions the MPS may already serve as a good approximation which captures dominant features of the distribution. 

\subsubsection{Adaptive Adjustment of Expressibility}

Performing optimizations for the two-site tensor instead of for each tensor individually, allows one to dynamically adjust the bond dimensions during the learning process. 
Since for realistic datasets the required bond dimensions are likely to be inhomogeneous, adjusting them dynamically allocates computational resources in an optimal manner. This situation will be illustrated clearly using the MNIST data set in Sec.~\ref{sec:mnist}, and in Fig.~\ref{fig:MPSbonddim}.  

Adjustment of the bond dimensions follows the distribution of singular values in~(\ref{eq:SVD}), which is related to the low entanglement inductive bias of the MPS representation. Adaptive adjustment of MPS is advantageous compared to most other generative models. Because in most cases, the architecture (which is the main limiting factor of the expressibility of the model) is fixed during the learning procedure, only the parameters are tuned. By adaptively tuning the bond dimensions, the representational power of MPS can grow as it gets more acquainted with the training data. In this sense, adaptive adjustment of expressibility is analogous to the structural learning of probabilistic graphical models, which is, however, a challenging task due to discreteness of the structural information. 

\subsubsection{Efficient Computation of Exact Gradients and Log-likelihood}
Another advantage of MPS compared to the standard energy based model is that training can be done with high efficiency. The two terms contributing to the gradient \eqref{eq:gradient:1} are analogous to the negative and positive phases in the training of energy based models~\cite{lecun2006tutorial}, where the visible variables are unclamped and clamped respectively. In the energy based models, such as RBM, typical evaluation of the first term requires approximated MCMC sampling ~\cite{RBMtraining}, or sophisticated mean-field approximations e.g. Thouless-Anderson-Palmer equations~\cite{Gabrie2015}. Fortunately, the normalization factor and its gradient 
can be calculated exactly and straightforwardly for MPS. 
The exact evaluation of gradients guarantees the associated stochastic gradient descent unbiased.

In addition to efficiency in computing gradients, the unbiased estimate of the log-likelihood and its gradients benefits significantly when compared with classic generative models such as RBM, where the gradients are approximated due to the intractability of partition function. 
First, with MPS we can optimize the NLL directly, while with RBM the approximate algorithms such as Contrastive Divergence (CD) is essentially optimizing a loss function other than NLL. This results in a fact that some region of configuration space could never be considered during training RBM and a subsequently poor performance on e.g. denoising and reconstruction.
Second, with MPS we can monitor the training process easily using exact NLL instead of other quantities such as reconstruction error or pseudo-lilelihood for RBM, which introduce bias to monitoring~\cite{Consistency}.


\subsubsection{Efficient Direct Sampling}

The approach introduced in Sec.~\ref{sec:Zipper} allows direct sampling from the learned probability distribution. This completely avoids the slowing mixing problem in the MCMC sampling of energy based models. 
MCMC randomly flip the 
bits and compare the probability ratios for accepting and rejecting the samples. However, the random walks in the state space can get stuck in a local minimum, which may bring unexpected fluctuations of long time correlation to the samples. 
Sometimes this raises issues to the samplings. As a concrete example, consider the case where all training samples are exactly memorized by both MPS and RBM. This is to say that NLL of both models are exactly $\ln|\mathcal {T}|$, and only training samples have finite probability in both models. While other samples, even with only one bit different, have zero probability. 
It is easy to check that our MPS model can generate samples which is identical to one of the training samples using approach introduced in Sec.~\ref{sec:Zipper}. However, RBM will not work at all in generating samples, as there is no direction that MCMC could follow for increasing the probability of samplings.

It is known that when graphical models have an appropriate structure (such as a chain or a tree), the inference can be done efficiently~\cite{wainwright2008graphical, PGM}, while these structural constraints also limit the application of graphical models with intractable partition functions. The MPS model, however, enjoys both the advantages of efficient direct sampling and a tractable partition function. 
The sampling algorithm is formally similar to the ones of autoregressive models~\cite{NADE, pixelRNN} though, being able to dynamically adjust its expressibility makes the MPS a more flexible generative model.

Unlike GAN~\cite{GAN} or VAE~\cite{VAE}, the MPS can explicitly gives tractable probability, which may enable more unsupervised learning tasks.
Moreover, the sampling in MPS works with arbitrary prior information of samples, such as fixed bits, which supports applications like image reconstruction and denoising. We note that this offers an advantage over the popular GAN, which easily maps a random vector in the latent space to the image space, but having difficulties in the reverse direction --- mapping a vector in the images space to the latent space as a prior information to sampling.



\section{Applications \label{sec:expts}}
In this section, to demonstrate the ability and features of the MPS generative modeling, we apply it to several standard datasets.
As a proof of principle, we first apply our method to the toy data set of \textit{Bars and Stripes}, where some properties of our model can be characterized analytically. 
Then we train MPS as an associative memory to learn random binary patterns to study properties such as capacity and length dependences. Finally we test our model on the
 \textit{Modified National Institute of Standards and Technology database} (MNIST) to illustrate its generalization ability for generating and reconstructing images of handwritten digits.~\footnote{The code of these experiments have been posted at \url{https://github.com/congzlwag/UnsupGenModbyMPS}.}

\subsection{Bars and Stripes \label{sec:BS}}
\begin{figure}[t!]
\centering
\subfloat[]{\label{fig:BSgrida}
\includegraphics[width=0.39\linewidth]{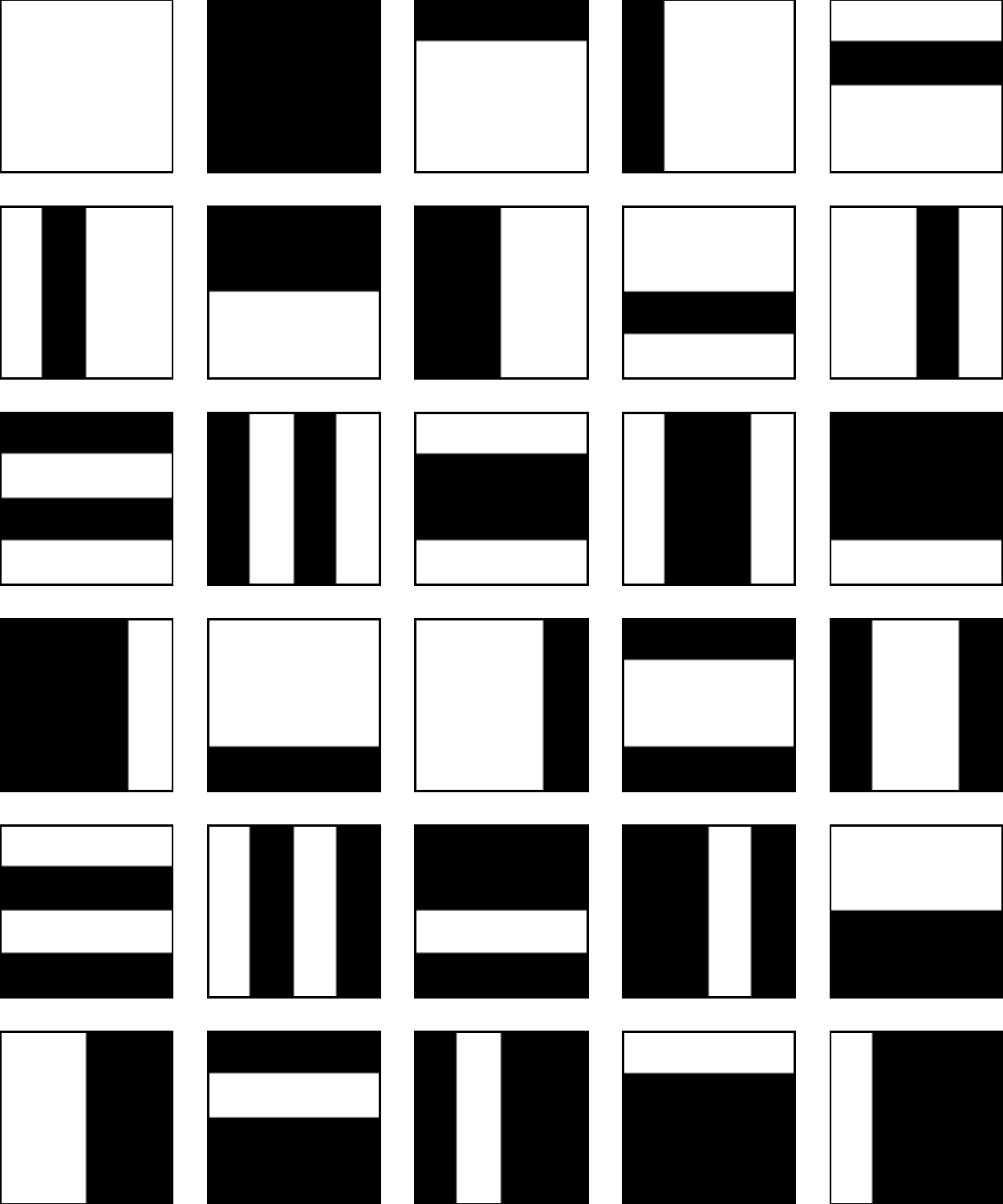}
}
\hspace{0.03\linewidth}
\subfloat[]{\label{fig:BSgridb}
\includegraphics[width=0.52\linewidth]{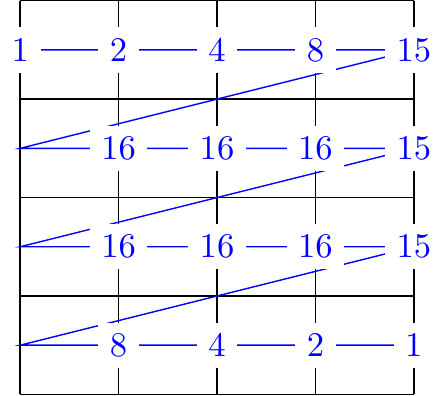}
}
\caption{(a) The Bars and Stripes dataset. (b) Ordering of the pixels when transforming the image into one dimensional vector. The numbers between pixels indicate the bond dimensions of the well-trained MPS.}
\label{fig:BSgrid}
\end{figure}

\textit{Bars and Stripes} (BS)~\cite{mackay2003information} is a data set containing $4\times 4$ binary images. Each image has either four-pixel-length vertical bars or horizontal stripes, but not both. In total there are $30$ different images in the dataset out of all $2^{16}$ possible ones, as shown in Fig.~\ref{fig:BSgrida}. These images appear with equal probability in the dataset.
This toy problem allows a detailed analysis and reveals key characteristics of the MPS probabilistic model. 

To use MPS for modeling, we unfold the $4\times 4$ images into one dimensional vectors as shown in Fig.~\ref{fig:BSgridb}. After being trained over $4$ loops of batch gradient descent training 
the cost function converges to its minimum value, which equals to the Shannon entropy of the BS dataset $\mathcal{S}=\ln (30)$, within an accuracy of $1\times 10^{-10}$. 
Here what the MPS has accomplished is memorizing the thirty images rigidly, 
by increasing the probability of the instances appeared in the dataset, and suppressing the probability of not-shown instances towards zero. We have checked that the result is insensitive to the choice of hyperparameters.

The bond dimensions of the learned MPS have been annotated in Fig.~\ref{fig:BSgridb}. It is clear that part of the symmetry of the data set has been preserved. For instance, the 180° rotation around the center or the transposition of the second and the third rows would change neither the data set nor the bond dimension distribution. Open boundary condition results in the decrease of bond dimensions at both ends. In fact when conducting SVD at bond $k$, there are at most $2^{\min(k,N-k)}$ non-zero singular values because the two parts linked by bond $k$ have their Hilbert spaces of dimension $2^{k}, 2^{N-k}$. In addition, the turnings bonds have slightly smaller bond dimension ($\mathcal{D}_{4}=\mathcal{D}_{8}=\mathcal{D}_{12}=15$) than others inside the second row and the third row, which can be explained qualitatively as 
these bonds carrying less entanglement than the bonds in the bulk. 

One can directly write down the exact ``quantum wave function'' of the BS dataset,
which has finite and uniform amplitudes for the training images and zero amplitude for other images. For division on  each bond, one can construct the reduced density matrix whose eigenvalues are the square of the singular values. Analyzed in this way, it is confirmed that the trained MPS achieves the minimal number of required bond dimension to exactly describe the BS dataset. 

We have generated $N_\mathrm{s}=10^6$ independent samples from the learned MPS. All these samples are training images shown in Fig.~\ref{fig:BSgrida}. Carrying out likelihood ratio test~\cite{wilks1938}, we got the log-likelihood ratio statistic 
$G^2=2N_\mathrm{s} D_\mathrm{KL}(\{\frac{n_j}{N_\mathrm{s}}\}||\{p_j\})=22.0$, equivalently $D_\mathrm{KL}(\{\frac{n_j}{N_\mathrm{s}}\}||\{p_j\})=1.10\times 10^{-5}$. 
The reason for adopting this statistic is that it is asymptotically $\chi^2$-distributed \cite{wilks1938}. The $p$-value of this test is $0.820$
, which indicates a high probability that the uniform distribution holds true for the sampling outcomes.

Note that $D_\mathrm{KL}(\{\frac{n_j}{N_\mathrm{s}}\}||\{p_j\})$ quantifies the deviation from the expected distribution to the sampling outcomes, so it reflects the performance of sampling method rather than merely the training performance.
In contrast to our model, for energy based models one typically has to resort to MCMC method for sampling new patterns. It suffers from slow mixing problem since various patterns in the BS dataset differs substantially and it requires many MCMC steps to obtain one independent pattern.

\subsection{Random patterns}

Capacity 
represents how much about data could be learned by the model.
Usually it is evaluated using randomly generated patterns as data. For the classic Hopfield model~\cite{Hopfield1982} with pairwise interactions given by Hebb's rule among $N\rightarrow\infty$ variables, it has been shown~\cite{Amit1985} that in the low-temperature region at the thermodynamic limit there is the retrieval phase where at most $|\mathcal{T}|_c=0.14N$ random binary patterns could be remembered.
In this sense, each sample generated by the model has a large overlap with one of the training pattern. 
If the number of patterns in the Hopfield model is larger than $|\mathcal{T}|_c$, the model would enter the spin glass state where samples generated by the model are not correlated with any training pattern.

Thanks to the tractable evaluation of the partition function ${Z}$ in MPS, we are able to evaluate exactly the likelihood of every training pattern. Thus the capability of the model can be easily characterized by the mean negative log-likelihood $\mathcal{L}$. 
In this section we focus on the behavior of $\mathcal{L}$ with varying number of training samples and varying system sizes.

\begin{figure}[t!]
\subfloat[]{\label{fig:random_patternsa}
\includegraphics[width=0.5\linewidth]{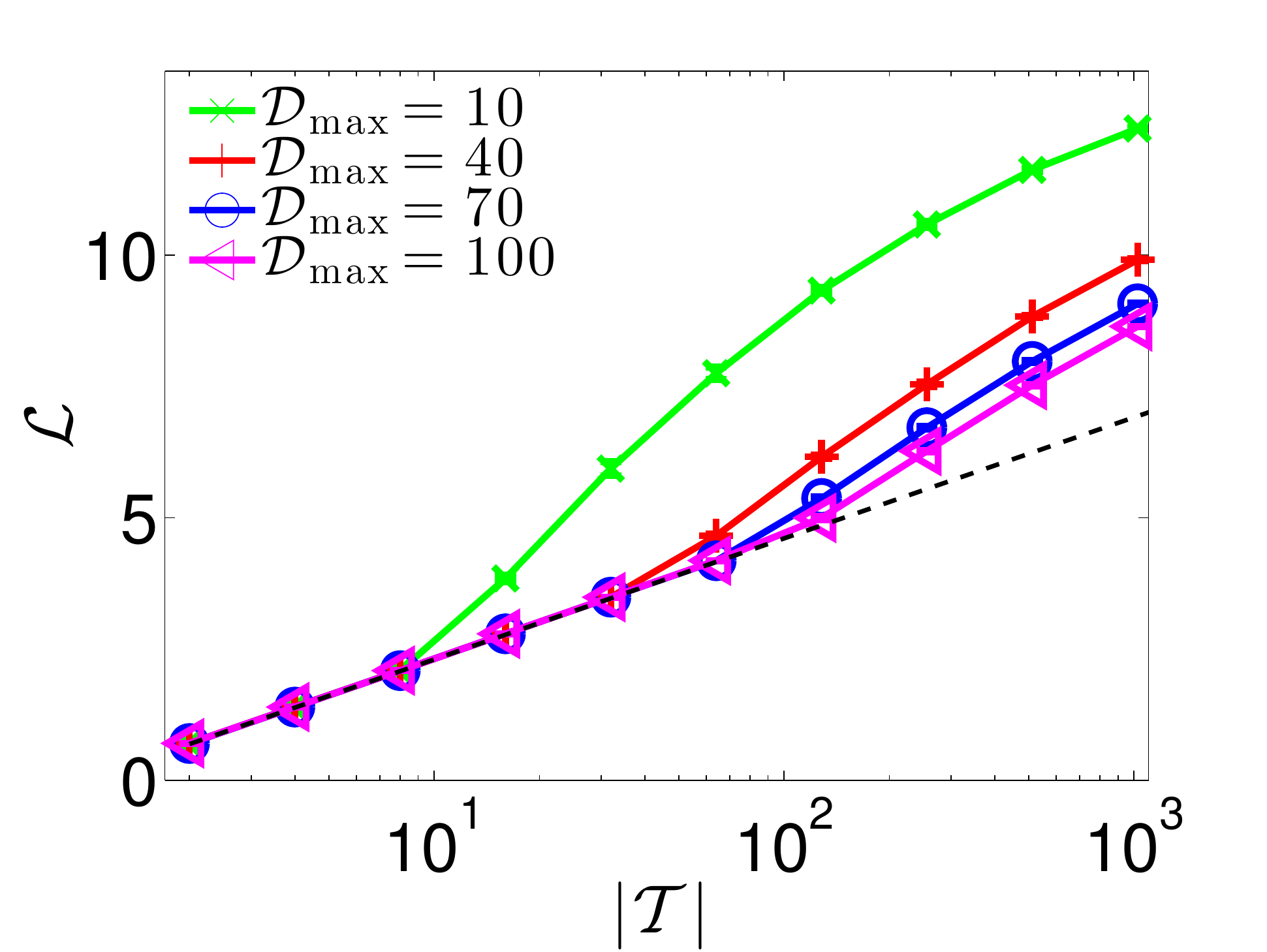}
}
\subfloat[]{\label{fig:random_patternsb}
\includegraphics[width=0.5\linewidth]{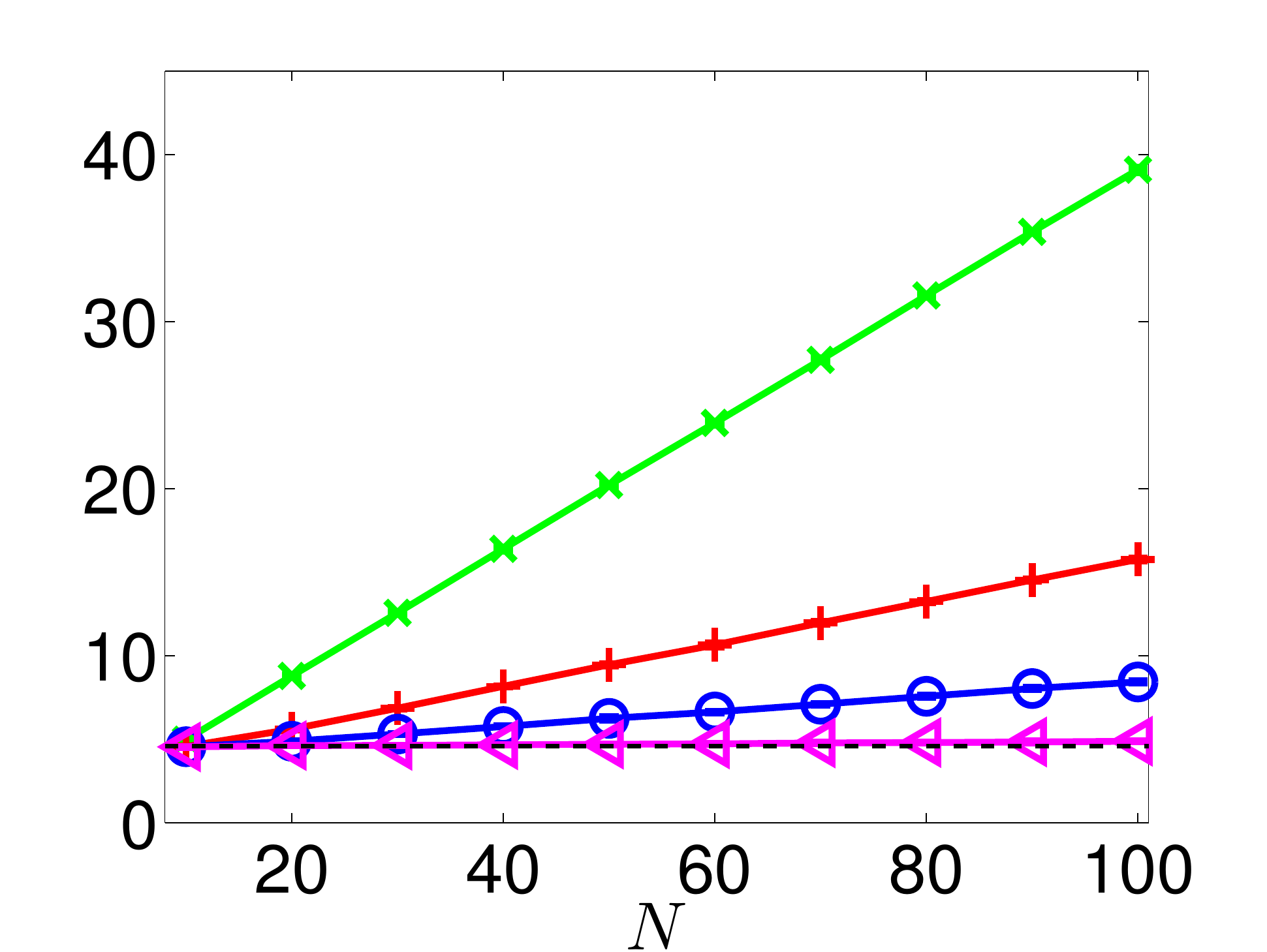}
}
\caption{
NLL averaged as a function of: (a) number of random patterns used for training, with system size $N=20$.  (b) 
system size $N$, trained using $|\mathcal{T}|=100$ random patterns. 
In both (a) and (b), different symbols correspond to different values of maximal bond dimension $\dmax$.
Each data point is averaged over $10$ random instances (i.e. sets of random patterns), error bars are also plotted, although they are much smaller than symbol size. The black dashed lines in figures denote $\mathcal{L}=\ln|\mathcal{T}|$.
}
\label{fig:random_patterns}
\end{figure}

\begin{figure}[t!]
\centering
\includegraphics[width=0.85\linewidth]{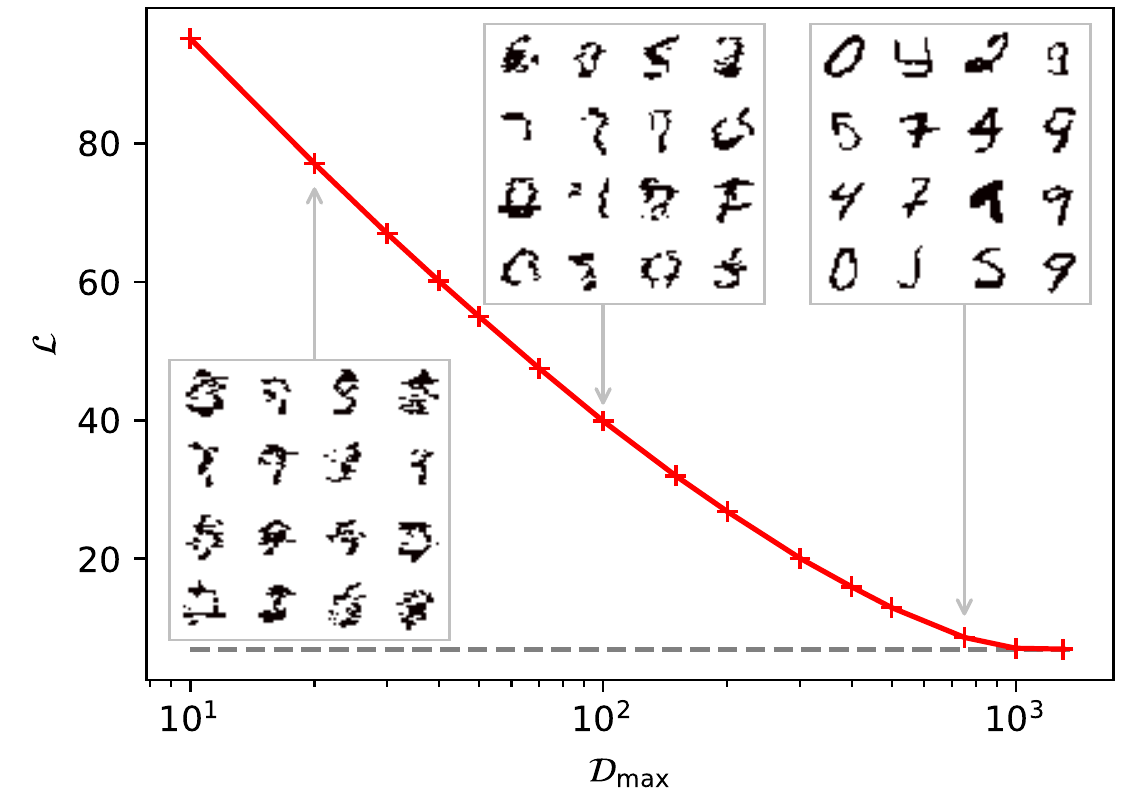}
\caption{ 
NLL averaged of a MPS trained using $|\mathcal{T}|=1000$ MNIST images of size $28\times 28$, with varying maximum bond dimension $\mathcal{D}_{\mathrm{max}}$.
 The horizontal dashed line indicates the Shannon entropy of the training set $\ln|\mathcal{T}|$, which is also the minimal value of $\mathcal{L}$.  The inset images are generated by the MPS' trained with different $\dmax$ (annoted by the arrows). }
\label{fig:mnist_m100}
\end{figure}

In Fig.~\ref{fig:random_patternsa} we plot $\mathcal{L}$ as a function of number of patterns used for training for several maximal bond dimension $\mathcal{D}_{\mathrm{max}}$. The figure shows that we obtain $\mathcal{L}=\ln |\mathcal{T}|$ for training set no larger than $\mathcal{D}_{\mathrm{max}}$.
As what has been shown in the previous section, this means that all training patterns are remembered exactly.
As the number of training patterns increases, MPS with a fixed $\mathcal D_{\mathrm{max}}$ will eventually fail in remembering exactly all the training patterns, resulting to $\mathcal{L} > \ln|\mathcal{T}|$. In this regime generations of the model usually deviate from training patterns (as illustrated in Fig.~\ref{fig:mnist_m100} on the MNIST dataset). 
We notice that with $|\mathcal{T}|$ increasing, the curves in the figure deviate from $\ln |\mathcal{T}|$ continuously. 
We note this is very different from the Hopfield model where the overlap between the generation and training samples changes abruptly due to the first order transition from the retrieval phase to spin glass phase.

Fig.~\ref{fig:random_patternsa} also shows that a larger $\dmax$ enables MPS to remember exactly more patterns, and produce smaller $\mathcal{L}$ with the number of patterns $|\mathcal{T}|$ fixed. This is quite natural because enlarging $\dmax$ amounts to the increase of parameter number of the model, hence enhances the capacity of the model. In principle if $\dmax=\infty$ our model has infinite capacity, since arbitrary quantum states can be decomposed into MPS~\cite{Schollwock2011}.
Clearly this is an advantage of our model over the Hopfield model and inverse Ising model~\cite{Roudi2009a}, whose maximal model capacity is proportional to system size.

Careful readers may complain that the inverse Ising model is not the correct model to compare with, because its variation with hidden variables, i.e. Boltzmann machines, do have infinite representation power. Indeed, increasing the bond dimensions in MPS has similar effects to increasing the number of hidden variables in other generative models.

In Fig.~\ref{fig:random_patternsb} we plot $\mathcal{L}$ as a function of system size $N$, trained on $|\mathcal{T}|=100$ random patterns. As shown in the figure that with $\dmax$ fixed $\mathcal{L}$ increases linearly with system size $N$, which indicates that our model gives worse memory capability with a larger system size. This is due to the fact that
keeping joint-distribution of variables becomes more and more difficult for MPS when the number of variables increases, especially for long-range correlated data. 
This is a drawback of our model when compared with fully pairwise-connected models such as the inverse Ising model, 
which is able to capture long-distance correlations of the training data easily.
Fortunately Fig.~\ref{fig:random_patternsb} also shows that the decay of memory capability with system size can be compensated by increasing $\dmax$. 

\subsection{MNIST dataset of handwritten digits}
\label{sec:mnist}

In this subsection we perform experiments on the MNIST dataset~\cite{MNIST}. In preparation we turn the grayscale images into binary numbers by threshold binarization and 
 flattened the images row by row into a vector. 
For the purpose of unsupervised generative modeling we do not need the labels of the digits. 
Here we further test the capacity of the MPS for this larger-scale and more meaningful dataset. Then we investigate its generalization ability via examining its performance on a separated test set, which is crucial for generative modeling.  

\subsubsection{Model Capacity}

Having chosen $|\mathcal{T}|=1000$ MNIST images, we train the MPS with different maximal bond dimensions $\dmax$, as shown in Fig.~\ref{fig:mnist_m100}. As $\dmax$ increases, the final $\mathcal{L}$ decreases to its minimum $\ln |\mathcal{T}|$, and the images generated become more and more clear. 
It is interesting that with a relatively small maximum bond dimension, e.g. $\dmax=100$, some crucial features show up, though some of the images were not as clear as the original ones. For instance the hooks and the loops that partly resembled to ``2'', ``3'' and ``9'' emerge.
These clear characters of handwritten digits illustrate that the MPS has learned many ``prototypes''. Similar feature-to-prototype transition in pattern recognitions could also be observed by using a many-body interaction in the Hopfield model, or equivalently using a higher-order rectified polynomial activation function in the deep neural networks~\cite{krotov2016dense}. It is remarkable that in our model this can be achieved by simply adjusting the maximum bond dimension of the MPS.

\begin{figure}[!t]
\centering
\includegraphics[width=0.9\linewidth]{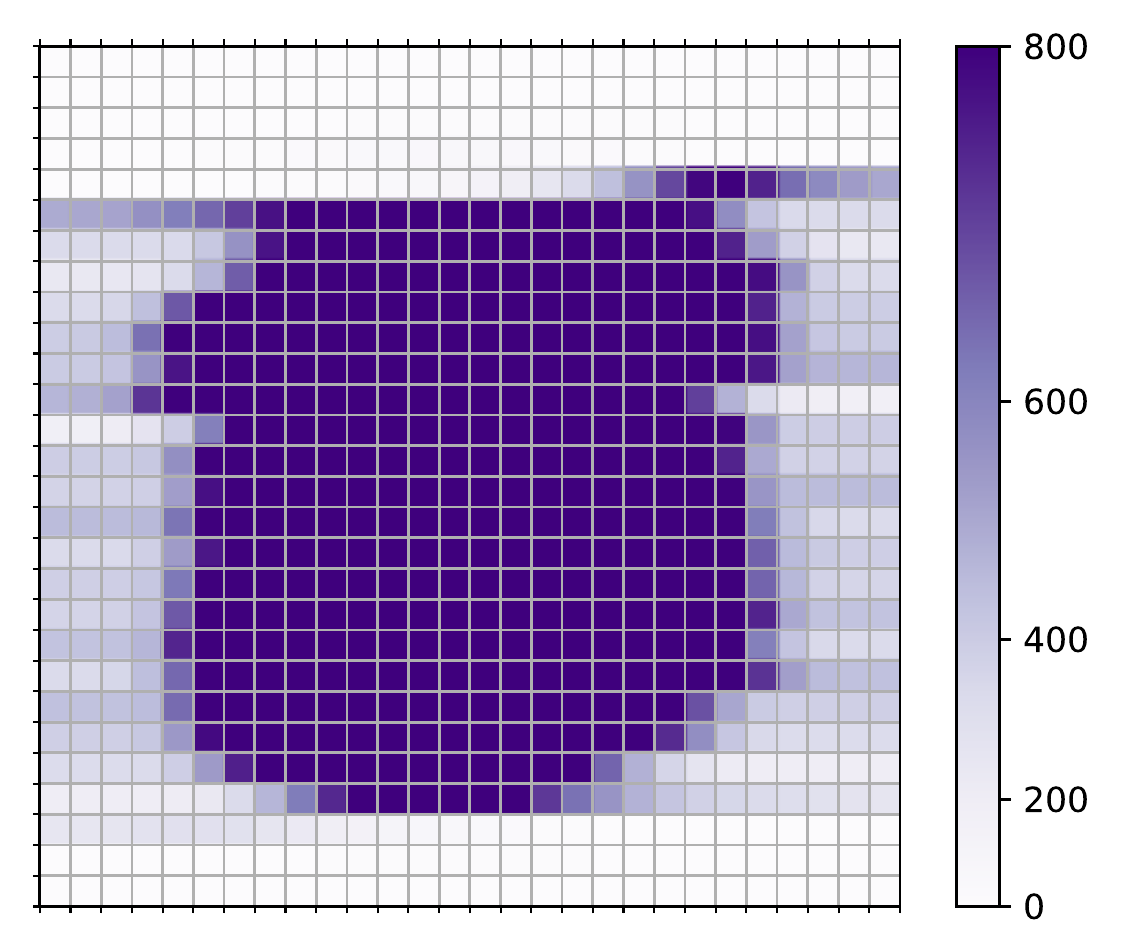}
\caption{Bond dimensions of the MPS trained with $|\mathcal{T}|=1000$ MNIST samples, constrained to $\dmax=800$. Final average NLL reaches
$16.8$. Each pixel in this figure corresponds to bond dimension of the right leg of the tensor associated to the identical coordinate in the original image.}
\label{fig:MPSbonddim}
\end{figure}

Next we train another model with the restriction of $\dmax=800$. 
The NLL on the training dataset reach $16.8$, and many bonds have reached maximal dimension $\dmax$. 
Fig.~\ref{fig:MPSbonddim} shows the distribution of bond dimensions. Large bond dimensions concentrated in the center of the image, where the variation of the pixels is complex. The bond dimensions around the top and bottom edge of the image remain small, because those pixels are always inactivated in the images. They carry no information and has no correlations with the remaining part of the image. Remarkably, although the pixels on the left and right edges are also white, they also have large bond dimensions because these bonds learn to mediate the correlations between the rows of the images. 

\begin{figure}[!b]
\centering
\subfloat[Generated]{\label{fig:MNIST-directa}
\includegraphics[width=0.45\linewidth]{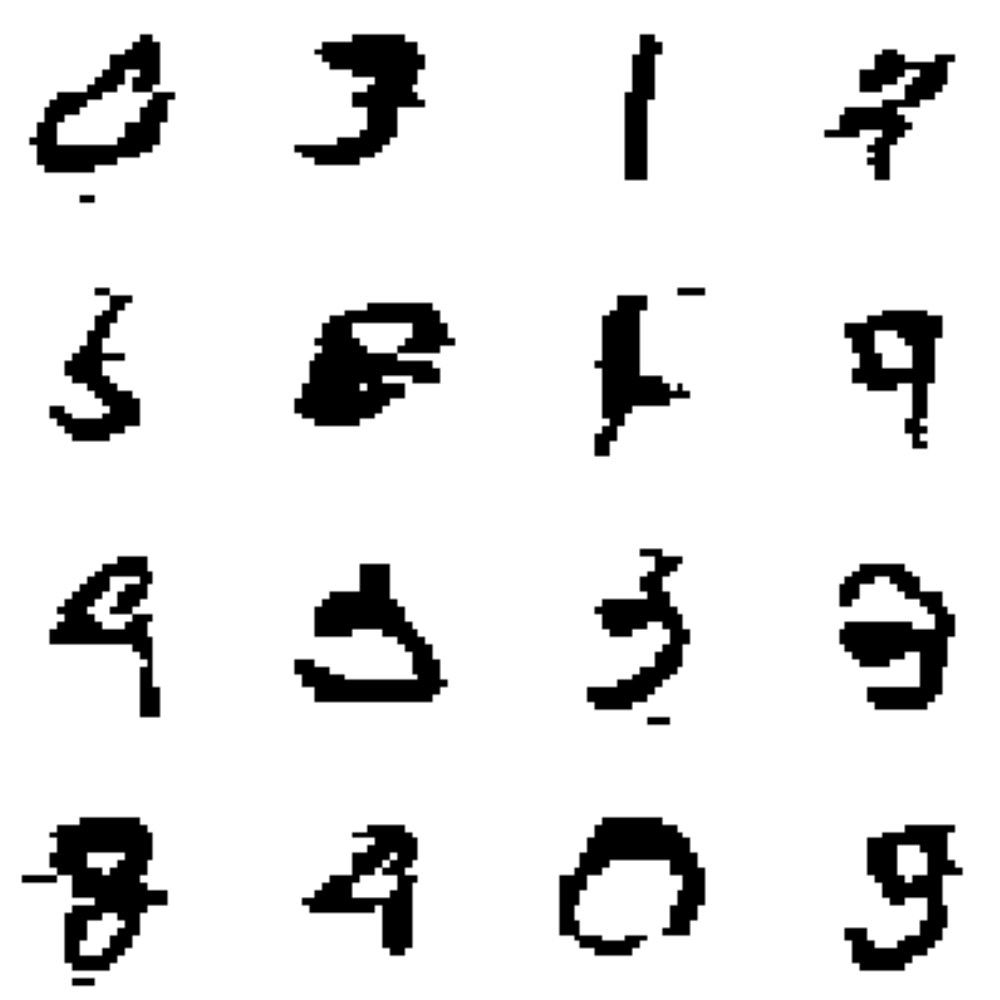}
}
\hspace{0.03\linewidth}
\subfloat[Original]{\label{fig:MNIST-directb}
\includegraphics[width=0.45\linewidth]{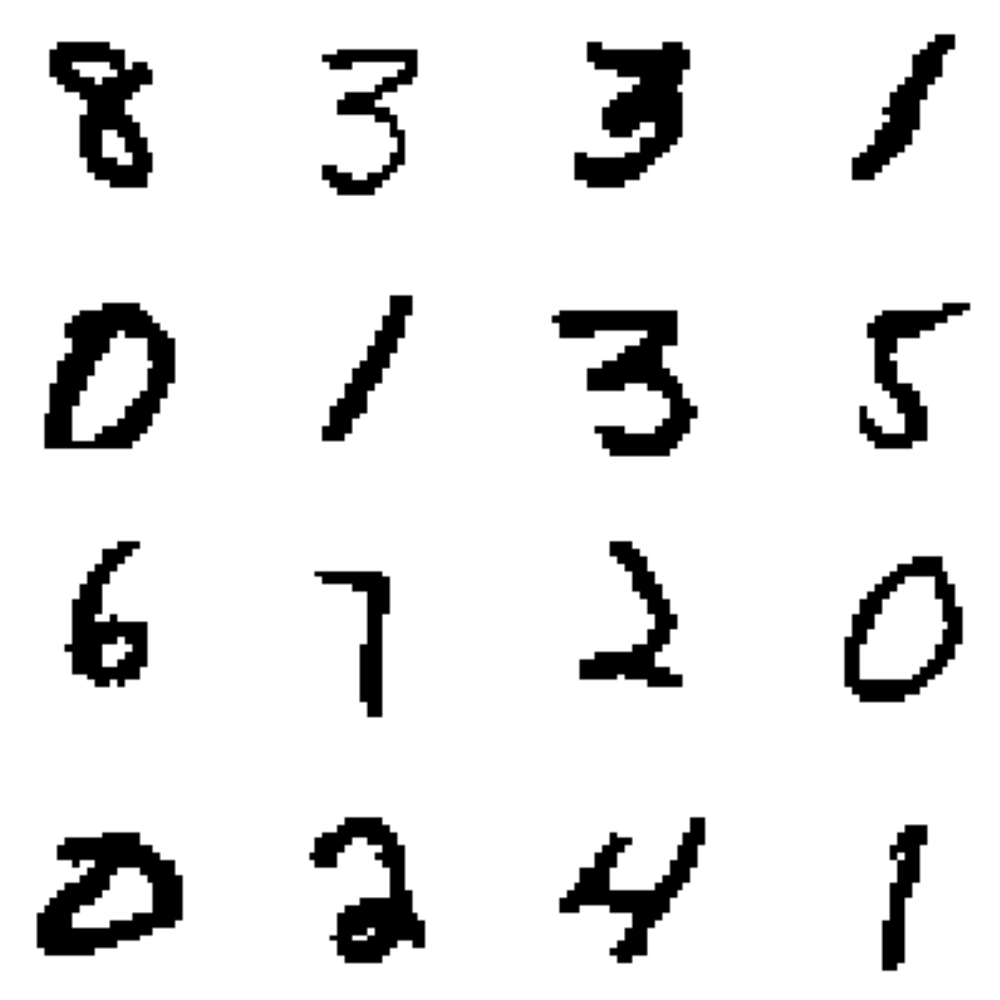}
}
\caption{(a) Images generated from the same MPS as in Fig.~\ref{fig:MPSbonddim}. (b) Original images randomly selected from the training set.}
\label{fig:MNIST-direct}
\end{figure}

The samples directly generated after training are shown in Fig.~\ref{fig:MNIST-directa}. We also show a few original samples from the training set in Fig.~\ref{fig:MNIST-directb} for comparison. Although many of the generated images cannot be recognized as digits, some aspects of the result are worth mentioning. Firstly, the MPS learned to leave margins blank
, which is the most obvious common feature in MNIST database. Secondly, the activated pixels compose pen strokes that can be extracted from the digits. Finally, a few of the samples could already be recognized as digits. 
Unlike the discriminative learning task carried out in \cite{schwab}, it seems we need to use much larger bond dimensions to achieve a good performance in the unsupervised task. 
We postulate the reason to be that in the classification task, local features of an image are sufficient for predicting the label.
Thus MPS is not required to remember longer-range correlation between pixels. For generative modeling, however, it is necessary  because learning the joint distribution from the data consists of (but not limited to) learning two-point correlations between pairs of variables that could be far from each other.

\begin{figure}[!t]
\centering


\subfloat[column reconstruction on training images]{\label{fig:MNIST-recona}
\includegraphics[width=0.45\linewidth]{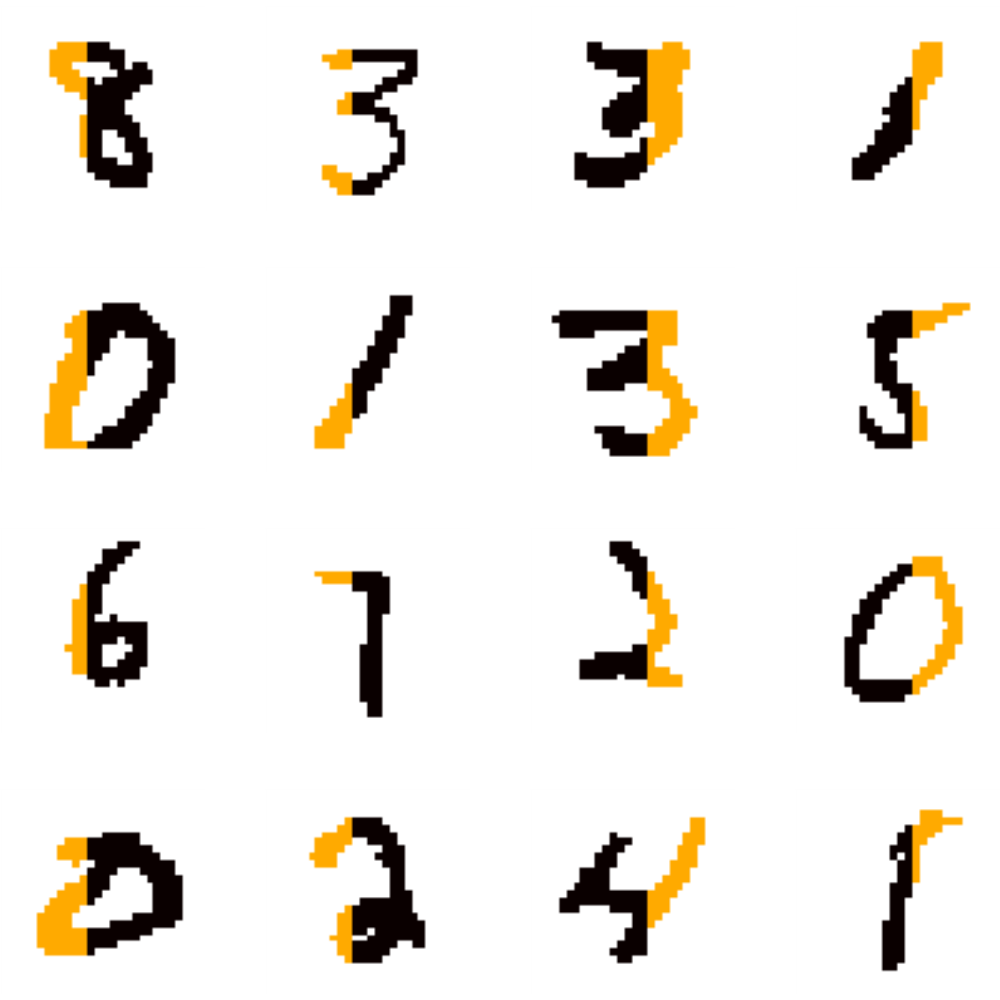}
}
\subfloat[row reconstruction on training images]{\label{fig:MNIST-reconb}
\includegraphics[width=0.45\linewidth]{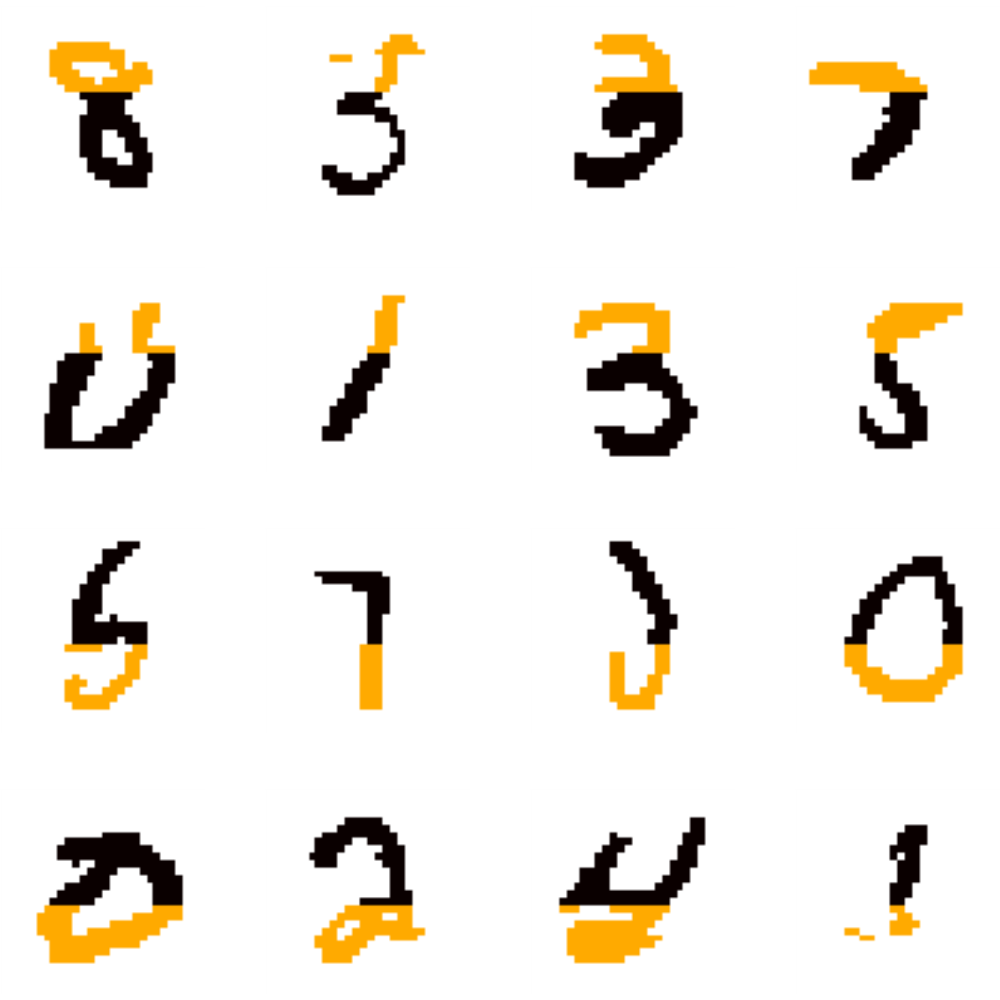}
}

\subfloat[column reconstruction on test images]{\label{fig:MNIST-reconc}
\includegraphics[width=0.45\linewidth]{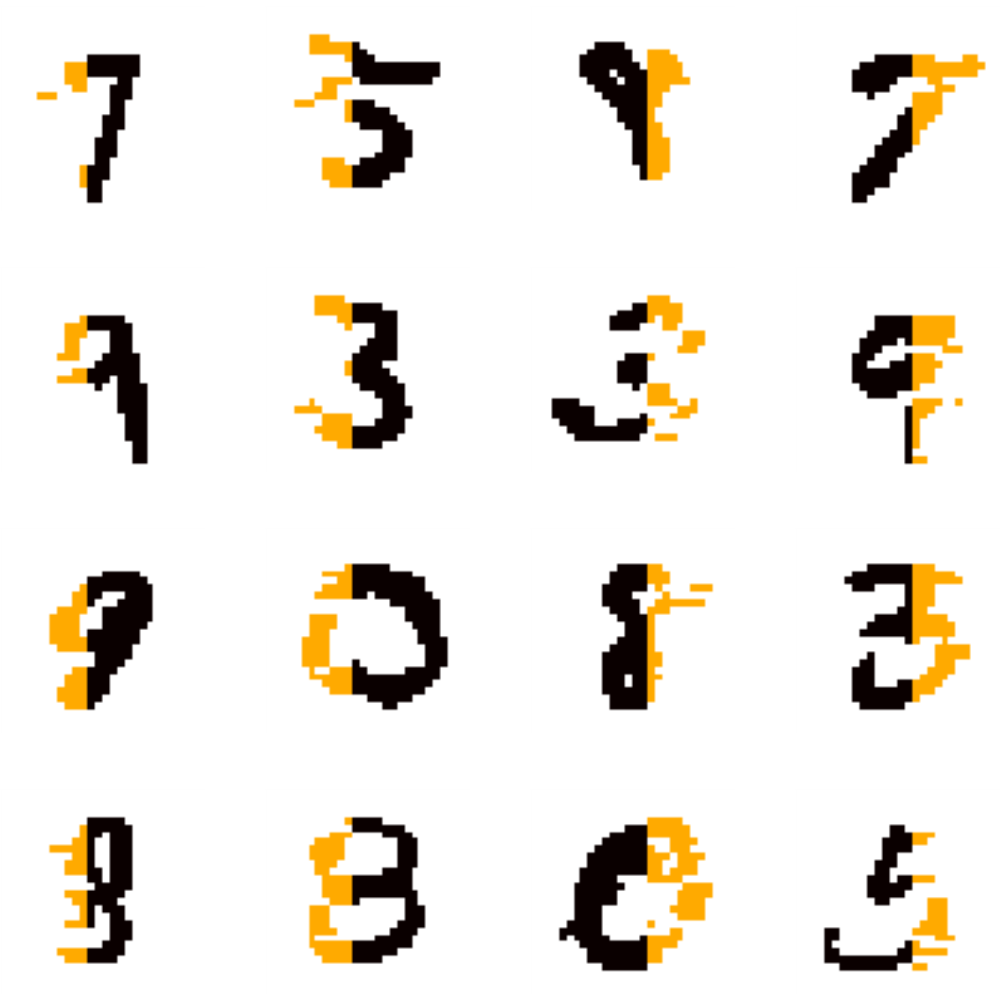}
}
\subfloat[row reconstruction on test images]{\label{fig:MNIST-recond}
\includegraphics[width=0.45\linewidth]{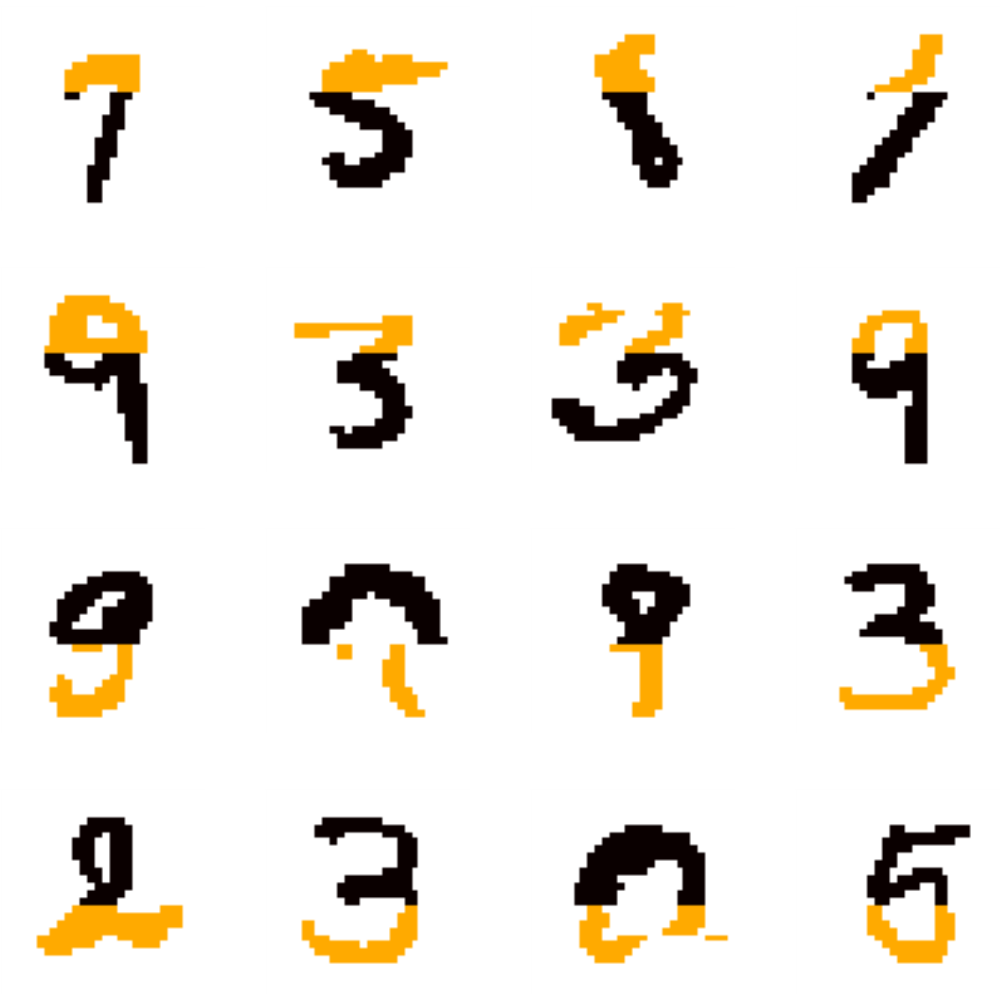}
}
\caption{
Image reconstruction from partial images by direct sampling with the same MPS as in Fig.~\ref{fig:MPSbonddim}. (a,b) Restoration of images in Fig.~\ref{fig:MNIST-directb} which are selected from the \emph{training} set . (c,d) Reconstruction of $16$ images chosen from the \emph{test} set. The test set contains images from the MNIST database that were not used for training. The given parts are in black and the reconstructed parts are in yellow. The reconstructed parts are:  $12$ columns from either (a,c) the left or the right, and (b,d) the top or the bottom. 
}
\label{fig:MNIST-recon}
\end{figure}

With the MPS restricted to $\dmax=800$ and trained with $1000$, we carry out image restoration experiments. As shown in Fig.~\ref{fig:MNIST-recon} we remove part of the images in Fig.~\ref{fig:MNIST-directb} and then reconstruct the removed pixels (in yellow) using conditional direct sampling. For column reconstruction, its performance is remarkable. The reconstructed images in Fig.~\ref{fig:MNIST-recona} are almost identical to the original ones in Fig.~\ref{fig:MNIST-directb}. On the other hand, for row reconstruction in Fig.~\ref{fig:MNIST-reconb}, it makes interesting but reasonable deviations. For instance, the rightmost in the first row, an ``1'' has been bent to ``7''.

\subsubsection{Generalization Ability}
In a glimpse of its generalization ability, we also tried reconstructing MNIST images other than the training images, as shown in Fig.~\ref{fig:MNIST-reconc}, \ref{fig:MNIST-recond}. These results indicate that the MPS has learned crucial features of the dataset, rather than merely memorizing the training instances. In fact, even as early as only $11$ loops trained, the MPS could perform column reconstruction with similar image quality, but its row reconstruction performance was much worse than that trained over $251$ loops. It is reflected that the MPS has learned about short range patterns within each row earlier than those with long range correlations between different rows, since the images have been flattened into a one dimensional vector row by row. 

\begin{figure}[h]
\centering
\includegraphics[width=1.0\linewidth]{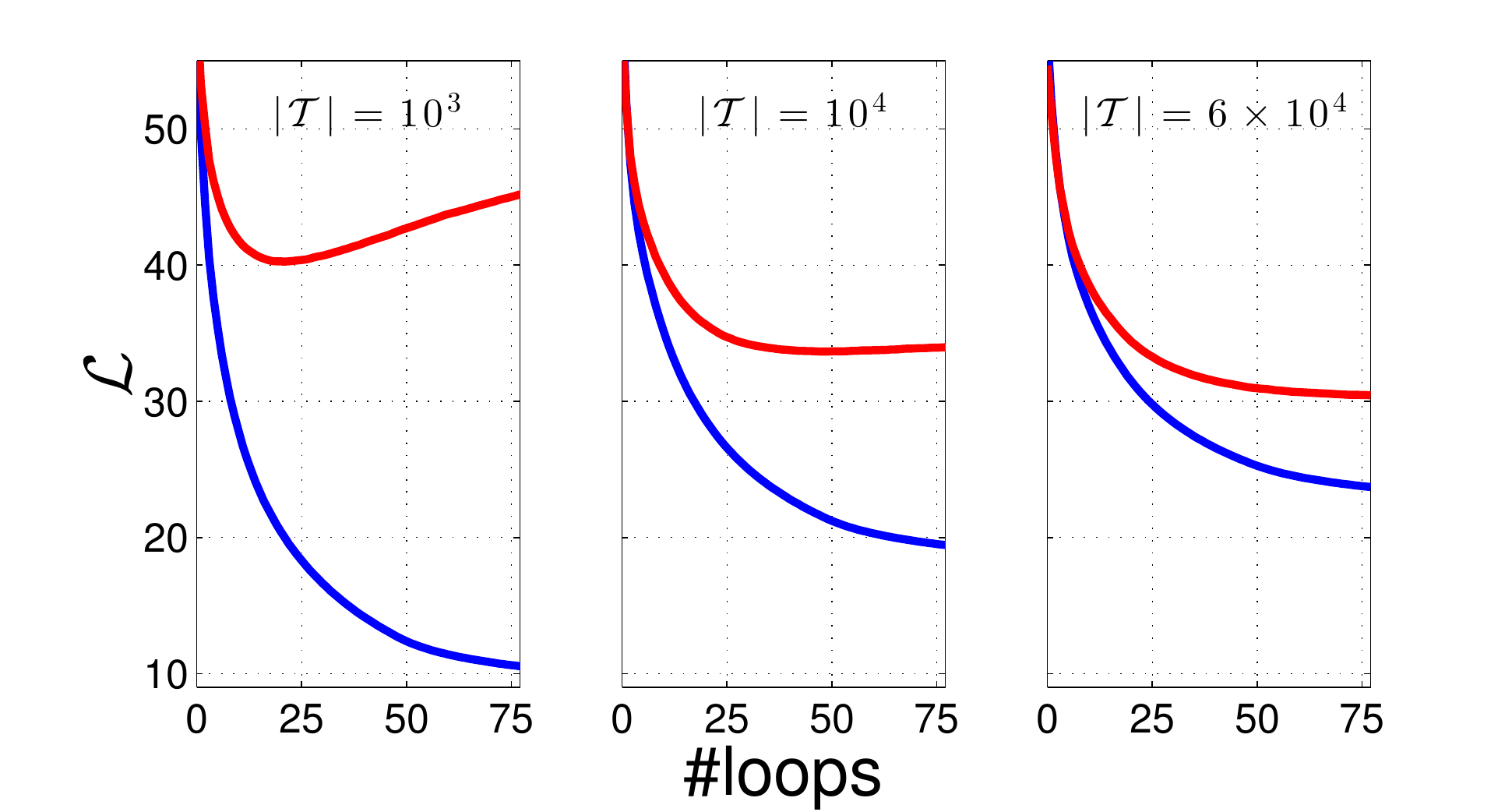}
\caption{ 
Evolution of the average negative log-likelihood $\mathcal{L}$ for both training images (blue, bottom lines) and $10^4$ test images (red, top lines) during training. From left to right, number of images in the training set $|\mathcal{T}|$ are $10^3, 10^4$, and  $6\times 10^4$ respectively.
\label{fig:generalization}
}
\end{figure}

To further illustrate our model's generalization ability, in Fig.~\ref{fig:generalization} we plotted $\mathcal{L}$ for the same $10^4$ test images after training on different numbers of images. To save computing time we worked on rescaled images of size $14\times 14$. The rescaling has also been adopted by past works, and it is shown that the classification on the rescaled images is still comparable with those obtained using other popular methods~\cite{schwab}. 

For different $|\mathcal T|$, $\mathcal{L}$ for training images always decrease monotonically to different minima, and with a fixed $\dmax$ it is easier for the MPS to fit fewer training images.
The $\mathcal{L}$ for test images, however, behaves quite differently: for $|\mathcal T|=10^{3}$, test $\mathcal{L}$ decreases to about $40.26$ then starts climbing quickly, while for $|\mathcal T|=10^4$ the test $\mathcal{L}$ decreases to $33.65$ then increases slowly to $34.18$. For $|\mathcal T|=6\times 10^4$, test $\mathcal{L}$ kept decreasing in $75$ loops. The behavior shown in Fig.~\ref{fig:generalization} is quite typical in machine learning problems. When training data is not enough, the model quickly overfits the training data, giving worse and worse generalization to the unseen test data. An extreme example is that if our model is able to decrease training $\mathcal{L}$ to $\ln|\mathcal{T}|$, i.e. completely overfits the training data, then all other images, even the images with only one pixel different from one of the training images, have zero probability in the model hence $\mathcal{L}=\infty$. We also observe that the best test NLL decreases as training set volume enlarges, which means the tendency of memorizing is constrained and that of generalization is enhanced. 

\begin{figure}[h]
\centering
\includegraphics[width=0.9\linewidth]{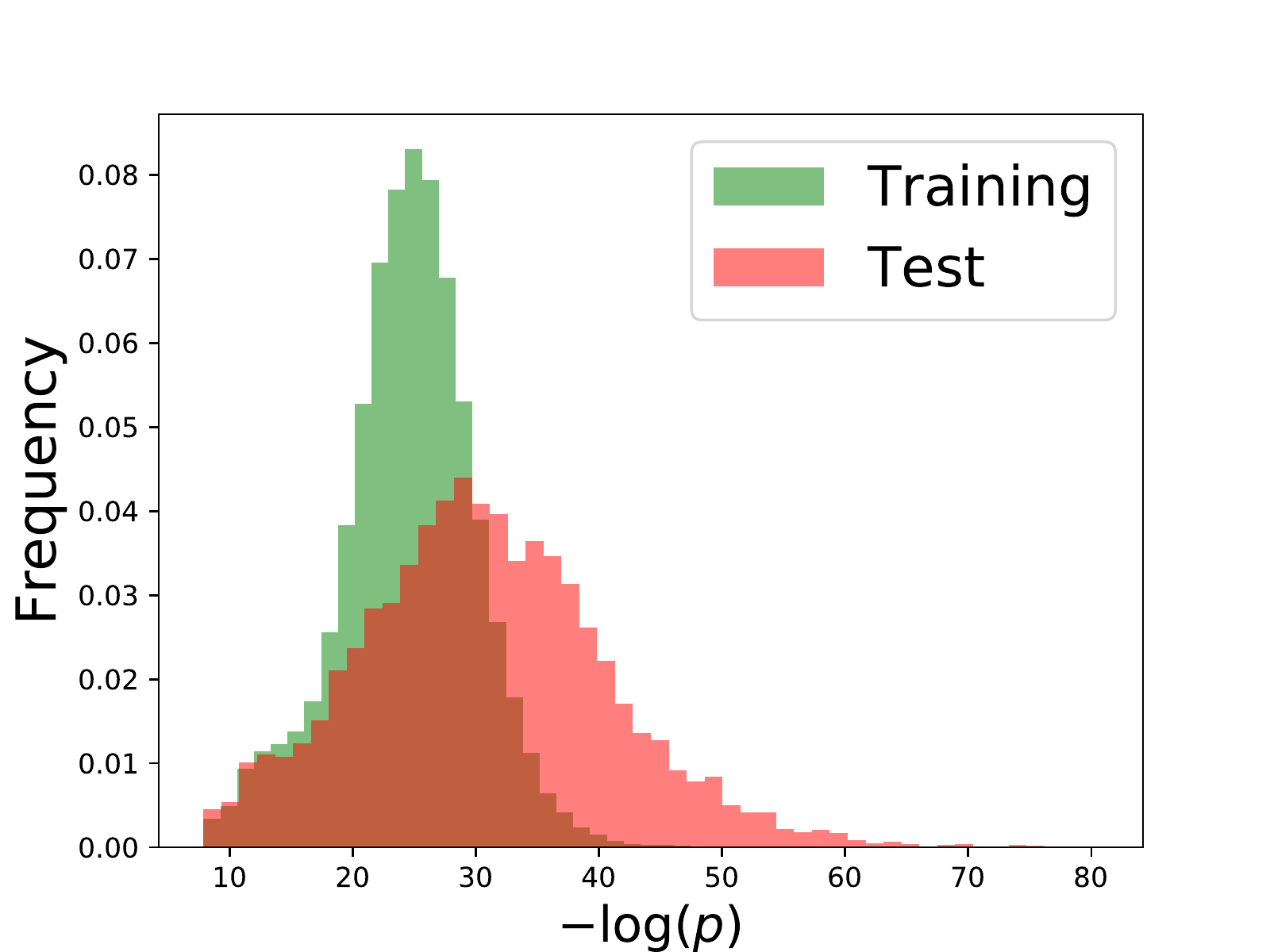}
\includegraphics[width=0.9\linewidth]{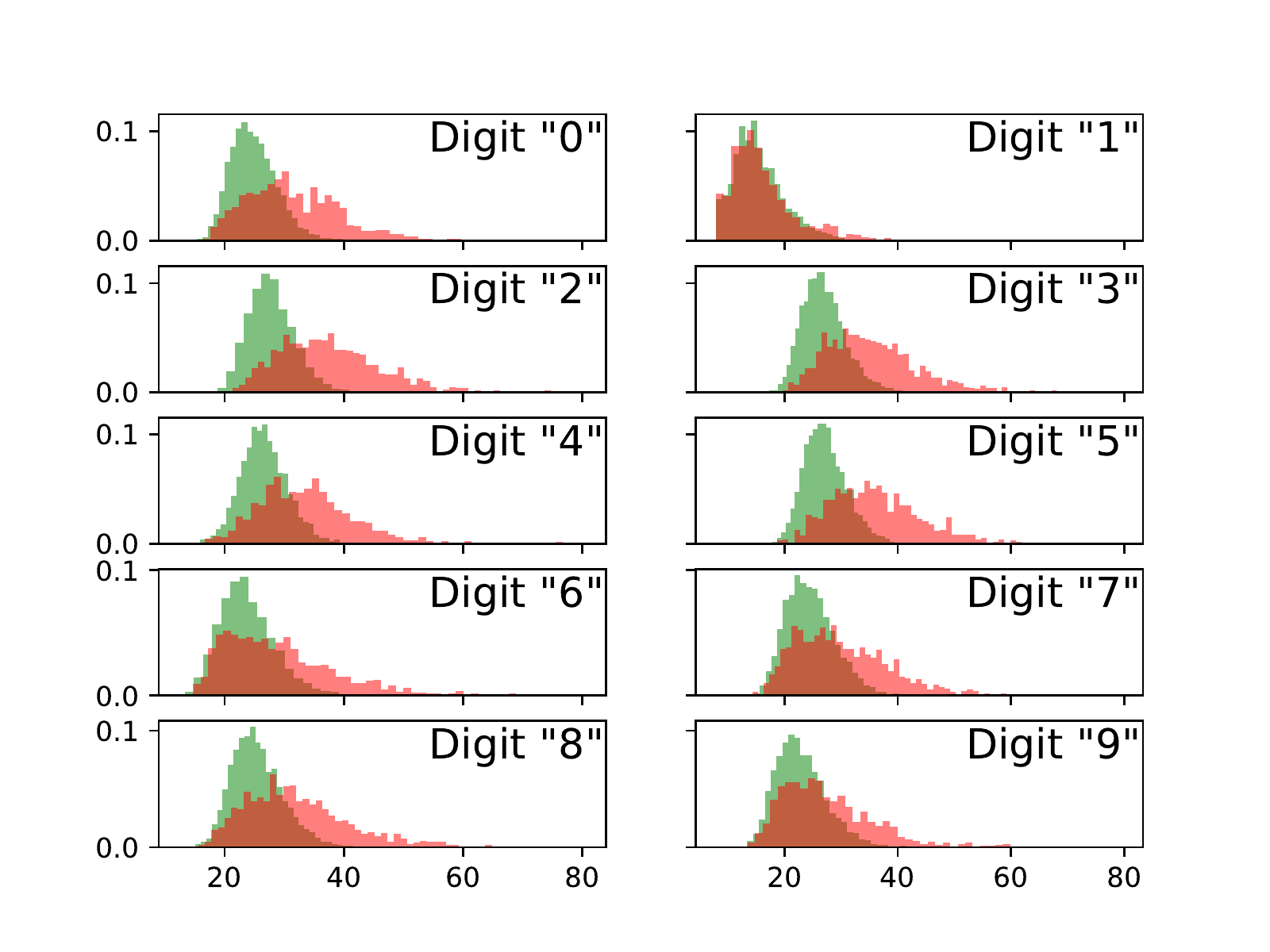}
\caption{(Top) Distribution of $-\ln p$ of $60000$ training images and $10000$ test images given by a trained MPS with $\mathcal{D}_{\textrm{max}}=500$. The training negative log likelihood $\mathcal{L}_{\textrm{train}}=24.2$, and the test $\mathcal{L}_{\textrm{test}}=30.3$. 
(Bottom) Distributions for each digit.
\label{fig:hist}
}
\end{figure}

The histograms of log-likelihoods for all training and test images are shown in Fig.~\ref{fig:hist}. Notice that if the model 
just memorized some of the images and ignored the others, the histograms would be bi-modal. It is not the case, as shown in the figure, where all distributions are centered around. This indicates that the model learns all images well rather than concentrates on some images while completely ignoring the others. In the bottom panel we show the detailed $\mathcal {L}$ histogram by categories. For some digits, such as ``1'' and ``9'', the difference between training and test log-likelihood distribution is insignificant, which suggests that the model has particularly great generalization ability to these images. 

\section{Summary and Outlook\label{sec:discussion}}

We have presented a tensor-network-based unsupervised model, which aims at modeling the probability distribution of samples in given unlabeled data. The probabilistic model is structured as a matrix product state, which brings several advantages as discussed in Sec.~\ref{sec:features}, such as adaptive and efficient learning and direct sampling. 

Since we use square of the TN states to represent probability, the sign is redundant  for probabilistic modeling besides the gauge freedom of MPS. It is likely that during the optimization MPS develops different signs for different configurations. The sign variation may unnecessarily increase the entanglement in MPS and therefore the bond dimensions~\cite{PhysRevLett.107.067202}. However, restricting the sign of MPS 
may also impair the expressibility of the model. One probable approach to obtain a low entanglement representation is adding a penalty term in the target function, for instance, a term proportional to R\'{e}nyi entanglement entropy as in our further work on quantum tomography~\cite{EfficientTomographyWithFidelityEstimation}. 
In light of these discussions, we would like point to future research on the differences and connections of MPS with non-negative
matrix entries~\cite{PhysRevLett.104.210502} and the probabilistic graphical models such as the hidden Markov model.


Binary data modeling links closely to quantum many-body systems with spin-$1/2$ constituents and could be straightforwardly generalized for higher dimensional data. 
One can also follow~\cite{schwab,stoudenmire2018learning} to use a local feature map to lift continuous variables to a spinor space for continuous data modeling.
The ability and efficiency of this approach may also depend on the specific way of performing the mapping, so in terms of continuous input there are still a lot to be explored on this algorithm. Moreover, for colored images one can encode the RGB values to three physical legs of each MPS tensor. 

Similar to using MPS for studying two-dimensional quantum lattice problems~\cite{2DDMRG}, modeling images with MPS faces the problem of introducing long range correlations for some neighboring pixels in two dimension. An obvious generalization of the present approach is to use more expressive TN with more complex structures. In particular, the projected entangled pair states (PEPS)~\cite{PEPS} is particularly suitable for images, because it takes care of correlation between pixels in two-dimension. Similar to the studies of quantum systems in 2D, however, this advantage of PEPS is partially compensated by the difficulty of contracting the network and the lose of convenient canonical forms. Exact contraction of a PEPS is $\#$P hard~\cite{PhysRevLett.98.140506}. Nevertheless, one can employ tensor renormalization group methods for approximated contraction of PEPS~\cite{PhysRevLett.99.120601, PhysRevLett.103.160601, PhysRevB.90.174201, PhysRevLett.115.180405}. Thus, it remains to be seen whether judicious combination of these techniques really brings a better performance to generative modeling. 

In the end, we would like to remark that perhaps the most exciting feature of quantum-inspired generative models is the possibility of being implemented by quantum devices~\cite{2017arXiv170809757P}, rather than merely being simulated in classical computers. In that way, neither the large bond dimension nor the high computational complexity of tensor contraction, would be a problem. The tensor network representation of probability may facilitate quantum generative modeling because some of the tensor network states can be prepared efficiently on a quantum computer~\cite{PhysRevLett.108.110502,huggins2018towards}. 

\begin{acknowledgments}
We thank Liang-Zhu Mu, Hong-Ye Hu, Song Cheng, Jing Chen, Wei Li, Zhengzhi Sun and E. Miles Stoudenmire for inspiring discussions. 
We also acknowledge suggestions from anonymous reviewers.
P.Z. acknowledges Swarm Club
workshop on ``Geometry, Complex Network and Machine
Learning'' sponsored by Kai Feng Foundation in 2016. L.W. is supported by the Ministry of Science and Technology of China under the Grant No. 2016YFA0300603 
and National Natural Science Foundation of China under the Grant No. 11774398. J.W. is supported by 
National Training Program of Innovation for Undergraduates of China. 
P.Z. is supported by Key Research Program of Frontier Sciences，CAS，Grant No. QYZDB-SSW-SYS032 and Project 11747601 of National Natural Science Foundation of China.
Part of the computation was carried out at the High Performance Computational Cluster of ITP, CAS.

\end{acknowledgments}




\bibliography{MyCollection.bib}
\appendix
\section{Canonical conditions for MPS and computation of the partition function}\label{sec:canonical}
The MPS representation has gauge degrees of freedom, which means that the state is invariant after inserting identity $I = MM^{-1}$ on each bond ($M$ can be different on each bond). Exploiting the gauge degrees of freedom, one can bring the MPS into its canonical form: for example, the tensor $A^{(k)}$
is called left-canonical if it satisfies $\sum_{v_{k}\in \{0, 1\}} \left( A^{(k)v_{k}} \right)^{\dagger}A^{(k)v_{k}} = I $. In diagrammatic notation, the left-canonical condition reads 
\begin{equation}
\label{left-can}
\tn{0.2}{3}
=
\tn{0.1}{4}
\end{equation}
The right-canonical condition is defined analogously. Canonicalization of each tensor can be done locally and only involves the single tensor at consideration~\cite{Schollwock2011, ORUS2014117}.

Each tensor in the MPS can be in a different canonical form. For example, given a specific site $k$, one can conduct gauge transformation to make all the tensors on the left, $\{A^{(i)}| i=1,2,\cdots, k-1\} $, left-canonical and tensors on the right, $\{A^{(i)}| i=k+1,k+2,\cdots, N\}$, right-canonical, while leaving $A^{(k)}$ neither left-canonical nor right-canonical. This is called mixed-canonical form of the MPS~\cite{Schollwock2011}. 
The normalization of the MPS is particularly easy to compute  in the canonical from. In the graphical notation, it reads
\begin{equation}
\label{eq:Norm}Z = 
\tn{0.35}{5}
=
\tn{0.154}{6}. 
\end{equation}
We note that even if the MPS is not in the canonical form, its normalization factor $Z$ can be still computed efficiently if one pays attention to the order of contraction~\cite{Schollwock2011, ORUS2014117}.





\section{DMRG-like Gradient Descent algorithm for learning}\label{sec:dmrg}
A standard way of minimization of the cost function \eqref{eq:NLL} is done by performing the gradient descent algorithm on the MPS tensor elements. Crucially, our method allows dynamical adjustment of the bond dimension during the optimization, thus being able to allocate resources to the spatial regions where correlations among the physical variables are stronger. 

Initially, we set the MPS with random tensors with small bond dimensions. For example, all the bond dimension are set to $\mathcal{D}_k=2$ 
except those on the boundaries~\footnote{Setting $\mathcal{D}_k=1$ for all bonds makes the bond dimension difficult to grow in the initial training phase. Since the rank of  two site tensor is $1\times2\times 2\times1$ and the number of nonzero singular value is at most $2$, which is likely to be truncated back to $\mathcal{D}_k=1$ with small cutoff.}.
We then carry out the canonicalization procedure so that all the tensors except the rightmost one $A^{(N)}$ are left-canonical. Then, we sweep through the matrices back and forth to tune the elements of the tensors, i.e. the parameters of the MPS. The procedure is similar to the DMRG algorithm with two-site update where one optimizes two adjacent tensors at a time~\cite{DMRG}. 
At each step, we firstly merge two adjacent tensors into an order-$4$ tensor, 
\begin{align}
\tn{0.4}{7}
=
\tn{0.286}{8},
\label{eq:2sitetensor}
\end{align}
followed by adjusting its elements in order to decrease the cost function 
 $\mathcal{L} = \ln Z -\frac{1}{|\mathcal{T}|}\sum_{ \ensuremath{\boldsymbol{v}} \in \mathcal{T}}\ln |\Psi(\ensuremath{\boldsymbol{v}} )|^2 $.
It is straight forward to check that its gradient with respect to an element of the tensor (\ref{eq:2sitetensor}) reads
\begin{equation}
\label{eq:gradient}
\frac{\partial\mathcal{L}}{\partial A^{(k,k+1)w_kw_{k+1}}_{i_{k-1}i_{k+1}}} = \frac{ Z^{\prime}}{Z} -\frac{2}{|\mathcal{T}|} \sum_{\ensuremath{\boldsymbol{v}}\in \mathcal{T}}  \frac{\Psi^{\prime}(\ensuremath{\boldsymbol{v}} )}{\Psi(\ensuremath{\boldsymbol{v}})} , 
\end{equation}
where $\Psi^{\prime}(\ensuremath{\boldsymbol{v}} )$ denotes the derivative of the MPS with respect to the tensor (\ref{eq:2sitetensor}), and 
$Z^{\prime} = 2\sum_{\ensuremath{\boldsymbol{v}}\in \mathcal{V}} \Psi^{\prime}(\ensuremath{\boldsymbol{v}} )\Psi(\ensuremath{\boldsymbol{v}} ) $. In diagram language, they read 
\begin{align} 
\label{psiprime} \Psi^{\prime}(\ensuremath{\boldsymbol{v}} ) &=
\tn{0.6}{9}\\
\frac{Z^{\prime}}{2} &= 
\tn{0.6}{10} \nonumber \\
& =
\tn{0.3}{11}\label{eq:Nprime}
\end{align}
The direct vertical connections of $w_k, v_k$ and $w_{k+1}, v_{k+1}$ in (\ref{psiprime}) stand for Kronecker delta functions $\delta_{w_{k}v_{k}}$ and  $\delta_{w_{k+1}v_{k+1}}$ respectively, meaning that only those input data with pattern $v_{k}v_{k+1}$ contribute to the gradient with respect to the tensor elements $A^{(k,k+1)v_{k}v_{k+1}}$. Note that although  $Z$ and $Z'$ involve summations over an exponentially large number of terms, they are tractable in MPS via efficient contraction schemes~\cite{Schollwock2011}. In particular, if the MPS is in the mixed canonical form, the computation only involves local manipulations illustrated in (\ref{eq:Nprime}). 


Next, we carry out gradient descent to update the components of the merged tensor.
The update is flexible and is open to various gradient descent techniques. Firstly, stochastic gradient descent is considerable. Instead of averaging the gradient over the whole dataset, the second term of the gradient (\ref{eq:gradient}) can be estimated by a randomly chosen mini-batches of samples, where the size of the mini-batch $m_\mathrm{batch}$ plays a role of hyperparameter in the training. Secondly, on a specific contracted tensor one can conduct several steps of gradient descent. Note that although the local update of $A^{(k,k+1)}$ does not change its environment, the shifting of $A^{(k,k+1)}$ makes a difference between $n_\mathrm{des}$ steps of update with learning rate $\eta$ and one update step with $\eta'=n_\mathrm{des}\times\eta$.
Thirdly, especially when several steps are conducted on each contracted tensor, the learning rate (the ratio of the update to the gradient) can be adaptively tuned by  meta-algorithms that such as RMSProp and Adam~\cite{kingma2014adam}.

In practice it is observed that sometimes the gradients become very small while it is not in the vicinity of any local minimum of the landscape. In that case a plateau or a saddle point may have been encountered, and we simply increase the learning rate so that the norm of the update is a function of the dimensions of the contracted tensor. 

After updating the order-$4$ tensor~(\ref{eq:2sitetensor}), it is decomposed by unfolding the tensor to a matrix, subsequently applying \textit{singular value decomposition} (SVD), and finally unfolding obtained two matrices back to two order-$3$ tensors.
\begin{align}
\tn{0.286}{8}
&=
\tn{0.4}{12}\nonumber\\
&=
\tn{0.4}{13}\nonumber\\
&\approx
\tn{0.4}{7}, \label{eq:SVD}
\end{align}
where $U, V$ are unitary matrices and $\Lambda$ is a diagonal matrix  containing singular values on the diagonal. The number of non-vanishing singular values will generally increase compared to the original value in \Eq{eq:2sitetensor} because the MPS observes correlations in the data and try to capture them. We truncate those singular values whose ratios to the largest one are smaller than a prescribed hyperparameter cutoff $\epsilon_\mathrm{cut}$, along with their corresponding row vectors and column vectors deleted in $U$ and $V^\dagger$. 

If the next bond to train on is the $(k+1)$-th bond on the right, take $A^{(k)} = U$ so that it is left-canonical, and consequently $A^{(k+1)}= \Lambda V^\dagger$. While if the MPS is about to be trained on the $(k-1)$-th bond, analogously $A^{(k+1)}=V^\dagger$ will be right-canonical and $A^{(k)}=U\Lambda$. This keeps the MPS in mixed-canonical form. 

The whole training process consists of many loops. In each loop the training starts from the rightmost bond (between $A^{(N-1)}$ and $A^{(N)}$) and sweeps to the leftmost $A^{(1)}$, then back to the rightmost.

\end{document}